\documentclass[preprint,12pt]{elsarticle}

\usepackage{amssymb}

\usepackage{amsmath}

\usepackage{listings}

\usepackage{microtype}

\journal{}

\begin{document}

\begin{frontmatter}

\title{Benchmarking LLM-Guided Control-Plane Policies for Backend Fault Isolation in HAProxy}

\author[label1]{Aman Chauhan}

\affiliation[label1]{organization={Independent Researcher},
            city={San Jose},
            postcode={95134},
            state={CA},
            country={USA}}

\author[label2]{Vishnu Pendyala}

\affiliation[label2]{organization={San José State University},
            addressline={1 Washington Sq},
            city={San Jose},
            postcode={95192},
            state={CA},
            country={USA}}

\begin{abstract}

Static load balancers cannot mitigate a backend that is degraded rather than down: round-robin and least-connections keep routing traffic to a server returning HTTP 500s until an operator intervenes.
We ask whether a Large Language Model can replace the static routing policy itself, reading HAProxy and Prometheus telemetry every 10 seconds and isolating faulty servers through guardrailed calls to the HAProxy Data Plane API.
On a reproducible benchmark with a persistent structural fault built into roughly one-third of a heterogeneous fleet, we sweep 15 open-weight models across five families (0.35B to 35B total parameters; dense, mixture-of-experts, and efficient-sparse architectures), reasoning modes, fleet scales of 3 to 9 backends, and two routing algorithms, totaling 240 runs.
We find a capability threshold near 3B active parameters.
Below it, LLM policies are typically unreliable and sometimes worse than no policy; above it, every model, regardless of architecture, saturates near an 88\% reduction in client-perceived 5xx errors over the static baseline.
The threshold is approximate: Gemma 4 E2B clears it with 2B active parameters, while the dense 3B Granite 4.0 Micro does not.
The availability gain has costs.
Draining concentrates load onto surviving servers, inflating tail latency 2.6 to 2.8 times, and enabling reasoning multiplies token spend roughly tenfold, overrunning the control interval and degrading effectiveness.
The efficient operating point is a supra-threshold model in its cheapest non-reasoning mode, wrapped inside deterministic guardrails.

\end{abstract}

\begin{graphicalabstract}

\includegraphics[width=\linewidth]{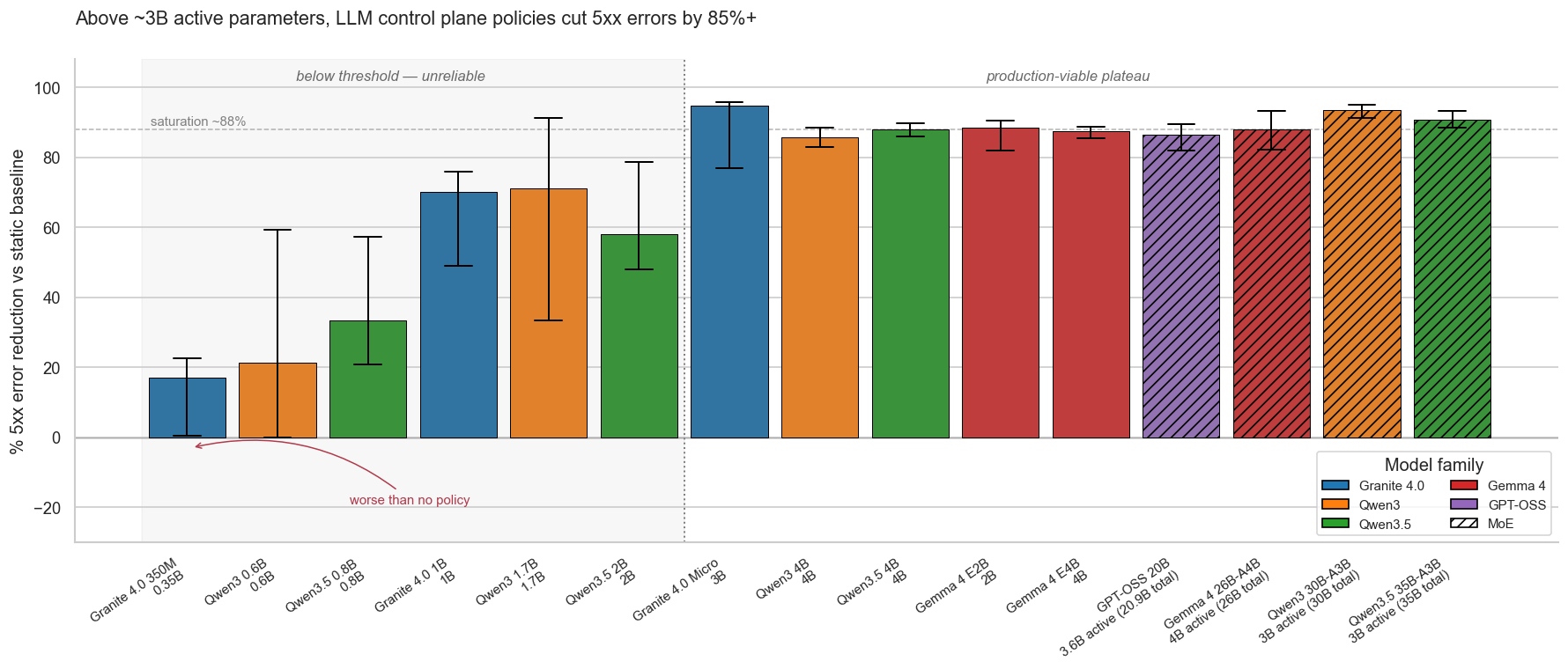}

\end{graphicalabstract}

\begin{highlights}

\item LLM as real-time HAProxy control plane policy reading raw telemetry.

\item Sweep of 15 open-weight models, 5 families, 0.35B-4B active params, 3-9 backends.

\item Capability threshold near 3B active params; $\sim$88\% 5xx reduction plateau.

\item Reasoning inflates cadence beyond 10s, cutting completed steps and effectiveness.

\end{highlights}


\end{frontmatter}

\section{Introduction}

Modern backend fleets are heterogeneous by default: servers differ in hardware generation, deployment history, and accumulated load, so even under steady traffic, a fraction of backends will return elevated 5xx errors.
Static load-balancing algorithms (round-robin and least-connections) are blind to application-layer faults.
They continue distributing traffic proportionally to a backend that returns HTTP 500s, propagating errors to every client whose request lands there until an operator intervenes or the server fails a health check.
For a consistent yet sub-total error rate (a backend that is degraded rather than down), these algorithms present no automatic mitigation, and the error burden falls directly on client-perceived availability.

Intent-Based Networking (IBN) separates operator goals from their implementation by structuring network management as a lifecycle: intent specification, where an operator expresses a high-level objective; translation, where that objective is decomposed into configuration artifacts; activation, where the artifacts are applied to the infrastructure; and assurance, where the live system is continuously monitored and corrected to maintain the specified intent.
Existing work has largely addressed the specification and translation phases, with closed-loop assurance at the control-plane level remaining underexplored.

This paper focuses on the assurance phase at the load balancer layer.
The system prompt that instructs the LLM corresponds to the intent specification: ``eliminate 5xx errors across the pool.''
The 10-second polling loop is the assurance cycle, continuously re-evaluating whether the intent is satisfied.
Tool calls to the HAProxy Data Plane APIs are the activation step that translates the LLM's decision into infrastructure state.

We instantiate a closed-loop LLM controller as a drop-in replacement for the static routing policy inside HAProxy.
Every 10 seconds, the controller

\begin{enumerate}
    \item queries Prometheus for per-backend metrics (5xx error rate, active sessions, weight, and up/down status),
    \item formats these as a compact text observation and passes them to a locally-served LLM via an OpenAI-compatible function-calling API, and
    \item executes the returned action plan against HAProxy through the Data Plane API and Runtime socket.
\end{enumerate}

The LLM's only available actions are adjusting server weights and draining servers.
A guardrail layer intercepts each action plan before execution, capping disruptive actions to at most one drain per interval, bounding weight deltas to 50 points per step, and limiting total actions to three per cycle.
Client-side latency metrics are withheld from the controller and reserved for post-hoc evaluation, thereby enforcing the same server-side-only observability as in a production deployment.

All experiments share a single fault model: a fixed subset of the fleet, every third instance, is permanently faulty by construction.
Servers whose instance index satisfies \texttt{idx \% 3 == 2} carry a 5\% per-request error rate baked into their startup configuration; all other backends run error-free.
The fault persists for the full 600-second run (the error rate never clears), so any re-enable action followed by a fresh drain reflects the LLM probing for recovery and re-quarantining upon observing continued errors.
The experiment gives the controller up to 60 decision intervals to identify and isolate the faulty backends.
There is no transient fault injection and no externally triggered recovery event.

We present the following contributions:

\begin{itemize}
  \item A systematic sweep of 15 open-weight models across five model families (Qwen3, Qwen3.5, Granite 4.0, Gemma 4, GPT-OSS), spanning 0.35B to 35B total parameters and 0.35B to 4B active parameters, each evaluated with reasoning enabled and/or disabled according to the model's supported modes, across two load-balancing algorithms and four fleet scales, totaling 224 LLM runs.
  \item Identification of a capability threshold at approximately 3B active parameters, below which models consistently fail to reliably isolate the faulty backends and above which reliable quarantine becomes achievable.
  \item A reproducible persistent-fault benchmark (a seeded 600-second load test with a deterministic heterogeneous backend fleet and a fixed fault pattern) that can serve as a reference point for future LLM control-plane evaluations.
  \item A comparison against static round-robin and least-connections baselines across the same four fleet scales (16 runs), establishing the conditions under which an LLM control plane reduces client-perceived 5xx errors.
\end{itemize}

\noindent Across the LLM sweep and the static baselines, the evaluation comprises 240 runs in total.

This paper discusses three research questions:

\begin{itemize}
  \item \textbf{RQ1}: When does an LLM control plane reduce client-perceived 5xx errors below the static load-balancer baseline?
  \item \textbf{RQ2}: How does LLM control-plane effectiveness degrade as the backend fleet grows?
  \item \textbf{RQ3}: Does reasoning effort improve LLM control-plane decisions enough to justify its token and latency cost?
\end{itemize}

\section{Related Work}

We organize prior work along the path from operator intent to live system operation: surveys mapping LLMs onto the IBN lifecycle; intent translation and extraction; closed-loop assurance; agentic control architectures; LLM adaptation for infrastructure operations; LLM-driven optimization; and the load-testing tooling used to validate such systems.
A consistent pattern runs through this literature: effort focuses on intent specification, translation, and orchestration-layer assurance, in which each decision cycle spans seconds to minutes and involves provisioning resources or generating policies.
Comparatively little work descends to the load-balancer data path, where a controller must react to noisy per-backend server-side metrics within a tight inference budget and without client-side visibility.
This paper occupies that gap.

\subsection{LLM Integration in Intent-Based Networking}

A survey examining LLM integration across all five IBN lifecycle phases finds that profiling and translation, addressed through conversational interfaces, RAG pipelines, and modular sub-intent architectures, are relatively mature.
At the same time, conflict resolution, activation, and assurance remain underexplored \cite{11186656}.
A complementary survey maps NLP, machine learning, and reinforcement learning to specific IBN functional layers, with NLP handling intent definition and translation to quality of service and routing policies, ML processing telemetry for anomaly detection, and RL driving dynamic routing and load-balancing decisions \cite{11125105}.
The first identifies the absence of public benchmarks and semantic drift in translation outputs as main obstacles \cite{11186656}, while the second adds scalability, data quality, vendor heterogeneity, and limited model interpretability \cite{11125105}; together, they establish runtime control, where intents must be upheld against live conditions as an open frontier.
This paper targets that frontier, treating an LLM as a runtime controller that observes live server-side metrics and continuously adjusts load-balancer behavior to uphold latency and availability objectives.

\subsection{Intent Translation and Extraction}

LLM-based intent extraction began with classification-only systems targeting 5G core management, using structured GPT-3.5 prompting aligned with 3GPP TS 28.312, producing human-readable labels but no machine-executable output \cite{10539172}.
Subsequent work introduced JSON-mediated intermediate representations: LLNet maps structured key-value pairs extracted from natural language to P4, OpenFlow, and eBPF API calls using a lightweight 3.8B-parameter model \cite{11003942}, and MicroIntent applies the same pattern to microservice SLO generation across the cloud-to-edge continuum \cite{11311632}.
NetIntent extends this to a full SDN lifecycle covering translation, validation, conflict recognition, and closed-loop assurance with benchmarks on 33 open-source LLMs showing that prompt design outweighs model scale \cite{11293797}.
All of these works treat intent fulfillment as a configuration-time task; none address controlling a live system from runtime metric telemetry.

\subsection{Closed-Loop Assurance}

Emergence decomposes intents as hierarchical policy trees executed by finite-state-machine-guided controllers, with LLM prompting automating the decomposition step \cite{10418089}; an extended version formalizes intent drift within a MAPE-K (monitor--analyze--plan--execute over shared knowledge) control loop, achieving 1-3x faster fulfillment than manual procedures \cite{11073734}.
Composite architectures combine BERT-based named-entity recognition with deep reinforcement learning for conflict resolution and multi-agent DRL for assurance, demonstrating full intent acceptance on a Healthcare 4.0 use case \cite{10742582}.
IntentContinuum positions GPT-4o as a real-time decision-maker that triggers placement, scaling, or SDN rerouting actions in response to SLO violations, achieving 85\% intent satisfaction on Kubernetes \cite{11169635}; NetIntent similarly incorporates a closed-loop assurance module that generates synthetic test traffic to diagnose drift \cite{11293797}.
All of these systems operate at the orchestration layer; none descend to load-balancer weight control driven by per-backend server-side metrics, the operating regime of this paper.

\subsection{Agentic Architectures for Network Control}

Multi-agent IBN frameworks decompose intents across specialist agents: a hierarchical 6G system delegates to RAN and Core Network agents through ReAct cycles, outperforming monolithic and rule-based baselines across ultra-reliable low-latency (URLLC), enhanced mobile-broadband (eMBB), and massive machine-type (mMTC) communication contexts \cite{jiang2026agenticaiempoweredintentbased}, while a distributed MCP-based framework exposes domain-specific tools to an Intent Decomposition Agent for Kubernetes and 5G core provisioning \cite{11131150}.
Principled credit assignment between agents is addressed by LLM-MCA and LLM-TACA, which employ an LLM critic to decompose joint trajectories into per-agent reward values, outperforming QMIX and MAPPO.
However, inference latency limits real-time use \cite{10.5555/3709347.3743784}.
These frameworks target configuration and deployment tasks, in which each decision cycle spans seconds to minutes.
This paper evaluates a different control problem at a comparable cadence: a guardrailed single-agent loop that runs every 10 seconds and adjusts load-balancer behavior in response to live server-side fault signals, rather than provisioning resources based on an operator's expressed intent, to examine whether it can outperform static load-balancing under multi-server fault conditions.

\subsection{LLM Adaptation for Infrastructure Operations}

DevOpsBERT fine-tunes BERT on system logs and cluster traces for offline anomaly detection and predictive maintenance but does not close the loop with live infrastructure \cite{10760725}.
NetLLM bridges the modality gap more systematically through a multimodal encoder, task-specific network heads, and parameter-efficient fine-tuning, outperforming rule-based and deep learning baselines for viewport prediction, adaptive bitrate, and cluster scheduling \cite{10.1145/3651890.3672268}.
An OpenStack VNF automation framework demonstrates that LLMs can generate operational scripts from procedural documents using runtime error feedback.
However, VNF configuration accuracy peaks at 62.3\%, revealing shortcomings in task-specific reasoning \cite{11073607}.
These works invest in LLM adaptation to produce correct network artifacts; this paper instead asks whether in-context reasoning alone suffices for real-time control decisions without fine-tuning.

\subsection{LLM-Driven Optimization and Resource Allocation}

Few-shot-prompted reasoning models (GPT O3-mini-high, O1, DeepSeek R1) produce near-optimal economic dispatch schedules without explicit mathematical formulations, though accuracy degrades near demand extremes \cite{11259643}.
Composite approaches pair LLMs with classical engines: MOEA/D-LPOS combines a DQN with an LLM for adaptive operator selection in multi-objective optimization \cite{11308679}, and MoED positions GPT-4o as an orchestrator over a mixture-of-experts library for service-function-chain (SFC) deployment, reducing solution time from hours to milliseconds \cite{11297203}.
Context-aware incident escalation in a security operations center (SOC) shows that LLM-RAG semantic matching combined with RL scheduling automates analyst assignment from unstructured operational data, though validated only in simulation \cite{11101733}; parameter-efficient LLM adaptation has likewise shown strong generalization to cluster job scheduling \cite{10.1145/3651890.3672268}.
All of these optimize over stable, well-defined problem formulations; load-balancer control requires reactive decisions amid noisy metric signals within tight inference budgets.

\subsection{Traffic Generation and Load Testing Tools for Experimental Validation}
\label{sec:rw-k6}

Grafana k6 has become a prevalent load generation tool for cloud and networking experiments, noted for its programmable arrival-rate control and native Prometheus integration.
Prior work uses it for Kubernetes distribution benchmarking \cite{https://doi.org/10.1002/spe.70000}, production trace replay against GPU serverless platforms \cite{10.1007/978-3-031-99854-6_11}, SLA-tier fairness evaluation \cite{hebbar2025sla}, RL policy feedback on Amazon EKS \cite{10.1007/978-3-032-10344-4_3}, and MicroK8s scaling characterization with sub-millisecond latency reproducibility \cite{app152412991}.
This paper follows these precedents by generating a deterministic, seeded workload for its evaluation scenario using k6.

\section{Methodology}

This section specifies the experiment end-to-end.
We first define the fault scenario and the actions available to the controller, then the four factors we vary, the static baselines that anchor every comparison, the metrics, and finally the full run matrix with its reproducibility guarantees.

\subsection{Scenario and Action Space}

We use a single \emph{baseline} scenario throughout: steady load against a heterogeneous fleet with no runtime fault injection.
Heterogeneity is \emph{structural} rather than event-driven: the 5\% error rate on the designated servers is part of their startup configuration, not an event triggered mid-run.
Each backend follows a fixed performance profile indexed by $i$ under a repeating mod-3 pattern (Table~\ref{tab:profiles}).
Hence, the disparity is present from $t=0$ and persists for the entire run.
One class of servers is consistently \emph{slow} and lossy (${\sim}9$ ms base latency, 5\% error rate); the rest are \emph{fast} and clean (7--8 ms, no errors).
The slow class comprises the $i \equiv 2 \pmod 3$ servers, roughly one-third of the fleet (exactly $1/3$ at $N\!\in\!\{3,9\}$, between $2/7$ and $2/5$ otherwise).
We evaluate at fleet sizes $N \in \{3,5,7,9\}$ and scale offered load proportionally (10 virtual users per backend), holding per-server pressure constant as the fleet grows.

This framing is deliberate: the correct long-run action is not to react to a transient incident but to detect a \emph{persistent steady-state} imbalance and respond by de-weighting or draining the consistently worst servers, concentrating traffic on the healthy majority.
The controller's stated objective is to drive pool-wide 5xx errors to zero.

\begin{table}[t]
\centering
\caption{Backend performance profiles by server index~$i$.
Latency $\sim \text{base} + \mathcal{U}\{0,\text{jitter}\}$ ms; base increases by $\lfloor i/3\rfloor$ ms to avoid duplicates at scale.}
\label{tab:profiles}
\begin{tabular}{@{}llccc@{}}
\hline
Class & Index & Base (ms) & Jitter (ms) & Error \\
\hline
Fast & $i \equiv 1 \!\!\pmod 3$ & 7 & 2 & 0\% \\
Fast & $i \equiv 0 \!\!\pmod 3$ & 8 & 3 & 0\% \\
Slow & $i \equiv 2 \!\!\pmod 3$ & 9 & 4 & 5\% \\
\hline
\end{tabular}
\end{table}

The controller selects from four primitive actions, exposed to the model as tools over an OpenAI-compatible function-calling API (Table~\ref{tab:actions}).
Every plan passes through a server-side guardrail layer before execution, applied identically to all policies (LLM and static baselines alike) and enforced per step: at most three actions, at most one drain, and a weight change of at most $\pm 50$.
Violating actions are dropped (excess or surplus drains) or clamped (out-of-range weight deltas), and every intervention is logged.

\begin{table}[t]
\centering
\caption{Controller action space}
\label{tab:actions}
\begin{tabular}{@{}llp{0.5\linewidth}@{}}
\hline
Action & Args & Effect \\
\hline
\texttt{set\_weight} & $s,\ w\!\in\![1,100]$ & Set server $s$'s integer traffic weight. \\
\texttt{drain}       & $s$ & Gracefully isolate $s$; no new sessions. \\
\texttt{enable}      & $s$ & Return a drained server to active routing. \\
\texttt{no\_op}      & --- & Record that no intervention is warranted. \\
\hline
\end{tabular}
\end{table}

\subsection{Independent Variables}

We vary four factors: the controller model (family and scale), its reasoning mode, the fleet size, and the load-balancer scheduling algorithm.
All controllers share an identical observation interface, action space, and system prompt; only the decision-making model changes.
Table~\ref{tab:models} lists the models; their exact weight revisions (HuggingFace repository, commit, and quantization) are given in Table~\ref{tab:checkpoints} of Appendix~\ref{app:checkpoints}.

\subsubsection{Model Family and Scale}

Five open-weight families span $0.35$--$35$ B parameters and three architectures (dense, MoE, and Gemma~4's efficient-sparse E-series), letting us separate the effect of scale from that of architecture.
A \emph{static} policy that never intervenes provides the no-controller reference; it is run twice, once with and once without the backend error profile, to bracket the achievable error range.

\begin{table}[t]
\centering
\caption{Model checkpoints evaluated in the sweep.
Active parameters are engaged per forward pass; total parameters include all experts.
Architecture codes: D = dense, MoE = mixture of experts, Eff = efficient sparse (Gemma~4 E-series).
All models are served as BF16/F16 GGUF weights via an OpenAI-compatible \texttt{llama-server} function-calling endpoint with a fixed decode seed.}
\label{tab:models}
\small
\begin{tabular}{llrrll}
\hline
\textbf{Family} & \textbf{Model} & \textbf{Active (B)} & \textbf{Total (B)} & \textbf{Arch.} & \textbf{Reasoning} \\
\hline
Qwen3      & 0.6B        & 0.6  & 0.6  & D   & think/nothink \\
           & 1.7B        & 1.7  & 1.7  & D   & think/nothink \\
           & 4B          & 4.0  & 4.0  & D   & think/nothink \\
           & 30B-A3B     & 3.0  & 30.0 & MoE & think/nothink \\
\hline
Qwen3.5    & 0.8B        & 0.8  & 0.8  & D   & think/nothink \\
           & 2B          & 2.0  & 2.0  & D   & think/nothink \\
           & 4B          & 4.0  & 4.0  & D   & think/nothink \\
           & 35B-A3B     & 3.0  & 35.0 & MoE & think/nothink \\
\hline
Gemma~4    & E2B         & 2.0  & 5.0  & Eff & think/nothink \\
           & E4B         & 4.0  & 8.0  & Eff & think/nothink \\
           & 26B-A4B     & 4.0  & 26.0 & MoE & think/nothink \\
\hline
Granite~4.0 & 350M       & 0.35 & 0.35 & D   & none \\
           & 1B          & 1.0  & 1.0  & D   & none \\
           & Micro       & 3.0  & 3.0  & D   & none \\
\hline
GPT-OSS    & 20B         & 3.6  & 20.9 & MoE & low/medium/high \\
\hline
\end{tabular}
\end{table}

\subsubsection{Reasoning Mode}

Reasoning Mode is a within-model toggle at fixed weights, applied through the chat template rather than a separate checkpoint.
Qwen3, Qwen3.5, and Gemma~4 expose a binary \texttt{enable\_thinking} flag; GPT-OSS~20B exposes a three-level \texttt{reasoning\_effort}; Granite~4.0 has no reasoning support.
When thinking is enabled, the model's reasoning trace is captured separately from its final rationale.
For Qwen, the flag is paired with a vendor-recommended, mode-specific sampling regime (Table~\ref{tab:sampling}); we follow this for Qwen3, but hold both Qwen3.5 modes at the thinking setting, varying only the flag.
Granite runs near-greedily ($T{=}0.001$) and is effectively deterministic; the decode seed is held fixed across all runs.

\begin{table}[t]
\centering
\caption{Decoding parameters for each model family.
Qwen3 resamples temperature and top-$p$ between thinking modes per vendor guidance; Qwen3.5 holds both modes at the thinking setting ($T{=}1.0$), varying only the template flag (marked $\dagger$).
GPT-OSS and Granite disable top-$k$.}
\label{tab:sampling}
\small
\begin{tabular}{lllll}
\hline
\textbf{Family} & \textbf{Mode} & $T$ & \textbf{top-}$p$ & \textbf{top-}$k$ \\
\hline
Qwen3       & think          & 0.6   & 0.95 & 20 \\
            & nothink        & 0.7   & 0.80 & 20 \\
Qwen3.5     & think          & 1.0   & 0.95 & 20 \\
            & nothink$^\dagger$ & 1.0 & 0.95 & 20 \\
Gemma~4     & both           & 1.0   & 0.95 & 64 \\
GPT-OSS~20B & all            & 1.0   & 1.00 & 0  \\
Granite~4.0 & ---            & 0.001 & 1.00 & 0  \\
\hline
\end{tabular}
\end{table}

\subsubsection{Fleet Scale}

$N \in \{3,5,7,9\}$ sets the observation dimensionality and the size of the degraded subset.
Every third backend (instances $2,5,8,\dots$) carries an elevated baseline 5xx rate and added latency, giving $1,2,2,3$ faulty servers at $N{=}3,5,7,9$ (fractions $\tfrac13, \tfrac25, \tfrac27, \tfrac39$); the degraded count grows with $N$, but its share is not held constant.
Offered load scales as $10N$ virtual users ($30/50/70/90$) to keep per-server load approximately constant.
The controller is invoked every $10$ s over a $600$ s run, giving $60$ decision steps.

\subsubsection{Load-Balancer Algorithm}

Runs use both \texttt{leastconn} and \texttt{roundrobin}, with the active algorithm disclosed to the model in the system prompt.
The load-balancing algorithm is not a primary treatment; we cross it with every other factor to check that the findings hold across HAProxy scheduling policies.

\subsection{Baselines}

We isolate the controller's contribution from that of HAProxy's native scheduler by evaluating against \emph{static} baselines.
In every baseline, the controller loop runs on its standard 10 s cadence, observing the system, querying stats, and recording results.
Still, the static policy always returns an empty plan: all backend weights stay fixed at 100, and no drain or enable action is ever issued.
We run two baselines, one per load-balancing algorithm, across server counts $N\in\{3,5,7,9\}$ with offered load scaled proportionally at $10N$ concurrent users.

Backends are heterogeneous and deterministic: each instance derives its latency and error rate from its index, and a per-request, SHA-256-seeded RNG makes the 5xx draw reproducible.
Every third backend (\texttt{api-2}, \texttt{api-5}, and so on) is a \texttt{slow} server that returns HTTP 500 on 5\% of requests, while the remainder are error-free.
Because the error process is stationary and seeded, the structural 5xx signal is visible within the first observation window.
Under no intervention, the aggregate failure rate is fixed by construction at $(N_{\text{slow}}/N)\times 5\%$, which we refer to as the \emph{no-intervention floor} (Table~\ref{tab:floor}).

\begin{table}[h]
\centering
\caption{No-intervention floor by fleet size: the aggregate 5xx rate when no policy acts.}
\label{tab:floor}
\begin{tabular}{cccc}
\hline
$N$ & Users ($10N$) & Faulty backends & No-intervention floor \\
\hline
3 & 30 & 1 & 1.7\% \\
5 & 50 & 2 & 2.0\% \\
7 & 70 & 2 & 1.4\% \\
9 & 90 & 3 & 1.7\% \\
\hline
\end{tabular}
\end{table}

The first scheduling baseline, \texttt{roundrobin}, has HAProxy rotate strictly across all backends irrespective of observed health, so faulty servers receive an equal traffic share.
This scheduling baseline exposes the full no-intervention floor and serves as our control condition.
The second, \texttt{leastconn}, has HAProxy route each request to the backend with the fewest active connections, a form of passive load-awareness that receives no signal from the controller.
Since connections are closed per request (\texttt{option http-server-close}), this baseline tests whether native scheduling alone is enough to suppress the structural signal.

For both algorithms, we additionally run a no-error mode in which every backend is configured with a 0\% error rate, turning off the per-request error draw entirely.
These runs remove the structural fault and yield the latency distribution that an effective controller should approach.
We treat it as an upper-bound reference rather than a competing policy.

\subsection{Metrics}

We evaluate each run along five questions, computed from the controller step log, the k6 client trace, and the per-step token-usage record: did the controller find the problem, how fast did it act, did it commit to its decision, did the fix reduce client-perceived failures, and what did the decisions cost.

\textbf{Ground truth}
HAProxy slot names do not map deterministically to backend containers (DNS resolution order reshuffles the mapping at each startup), so the faulty set is recovered post hoc per run.
Faulty slots accumulate a strictly positive cumulative 5xx count over the run while healthy slots stay at exactly zero, so the faulty set is simply the slots with any 5xx at all; this labeling is exact in every run, and its per-run cardinality is verified against the expected every-third-instance count.

\textbf{Time-to-action and time-to-resolution}
Time-to-action is the offset from run start to the first drain (or weight set to zero) on any faulty server.
Time-to-resolution is the offset until every faulty server has been drained at least once, undefined if never fully covered.

\textbf{Fault-localization precision/recall/F1}
Localization is treated as a per-run set comparison over backend slots, with the ground truth $G$ being the structurally faulty set and the prediction $P$ being the set of servers ever drained (or weighted to zero).
We report micro-averaged scores (pooling counts across runs, which weights runs by $|G|$) and the exact-match rate ($P{=}G$); a run that never drains scores $F1{=}0$.

\textbf{Stability}
Each server's actions collapse into an effective drained-versus-active timeline whose transitions after the first are \emph{flips}.
From this we derive the flip count, the number of oscillating servers ($\ge$2 flips), faulty- versus healthy-slot flip rates (normalized by the slots in each group), and end-state correctness (does the final drained set equal $G$?).
Contrasting end-state correctness with the ever-drained exact match quantifies how much instability undoes otherwise-correct decisions.

\textbf{Client-perceived 5xx rate and effectiveness}
The client-side 5xx rate is the fraction of requests receiving a 5xx response over the full run, measured independently by the k6 load generator rather than by the controller's own observations.
For each run, we report absolute and percentage rate reductions relative to the matched static, error-on baseline, using the same scale and algorithm, along with the reduction in absolute failed-request counts.
Within-run sampling uncertainty is quantified with a 95\% Wilson score interval on each run's 5xx rate and a 95\% Newcombe hybrid-score interval on each reduction (a difference of two binomial proportions).

\textbf{Latency} Per-request client latency from the k6 trace is summarized as mean, p50, p95, and p99 with 95\% confidence intervals, and binned into 10 s windows aligned to each run's start to visualize the latency trajectory.

\textbf{Token usage (inference-cost proxy)}
The per-step usage record yields prompt, completion, total, and cached token counts, the number of conversational turns, and the number of actions emitted.
Served locally with no per-token pricing, these counts serve as a hardware-agnostic proxy for the computational burden each model places on the inference server under the fixed 10 s cadence; pooled over the run, they also separate thinking from non-thinking configurations at the same checkpoint.
From the same record we derive a cost-effectiveness ratio, failed requests avoided (the absolute 5xx reduction against the matched static baseline) per $1{,}000$ completion tokens spent over the run, as an error-reduction return on inference spend.

\subsection{Run Matrix and Reproducibility}

The sweep comprises exactly 240 runs, all on the baseline scenario (no runtime fault injection; the structural error profile is active except in the error-off static arm), 600 s each, with seed 42 and scale/users drawn from \{$(3,30)$, $(5,50)$, $(7,70)$, $(9,90)$\} and algorithm in \{\texttt{leastconn}, \texttt{roundrobin}\}: 16 static baseline runs (error-on/error-off $\times$ algorithm $\times$ scale; this is the only arm where the error mode is varied), 128 Qwen3/Qwen3.5 runs (8 sizes $\times$ think/nothink $\times$ algorithm $\times$ scale), 24 Granite runs (3 sizes $\times$ algorithm $\times$ scale), 48 Gemma-4 runs (3 variants $\times$ think/nothink $\times$ algorithm $\times$ scale), and 24 GPT-OSS-20B runs (low/medium/high $\times$ algorithm $\times$ scale).
A single seed (42) propagates into the backend per-request RNG, the load generator's per-virtual-user PRNG, and the LLM server's decoder, making latency draws, error draws, arrival timing, and (where the runtime allows) decoding reproducible.

Sweep orchestration is idempotent: it skips runs whose output already exists, supports resumption from a given point, removes partial output on failure, and waits for each run's k6 summary before advancing.
The remote model server starts once per policy, stays warm across the reasoning/algorithm/scale sub-grid, and restarts only on a policy change, to amortize model-load cost.
Each run records its full configuration, a metric summary with total wall time, per-step records (observation, plan, execution results, post-observation), and the full LLM input/output, together enabling step-by-step replay and audit.

\section{Experimental Setup}

This section describes the testbed that realizes the methodology: the end-to-end system architecture, the backend fleet and its fault model, the telemetry and control surfaces, the LLM serving stack, the system prompt, and the load generator.

\subsection{System Architecture}

Requests flow from the k6 load generator through HAProxy to a Flask backend fleet scaled via container replication; HAProxy exposes a Prometheus metrics exporter, a Data Plane API for reads (backend/server discovery), and an admin stats socket for writes (weight, drain, enable).
A controller process reads Prometheus and HAProxy state, decides via a pluggable policy, and acts through those two surfaces.
Prometheus (v3.8.1, 1 s scrape interval) scrapes HAProxy and the backend fleet, discovering targets by DNS service discovery so the scrape set tracks the fleet at any scale.
\texttt{llama.cpp}'s OpenAI-compatible server is used to serve the LLM, run either as a local subprocess or, for the full sweep, remotely over an SSH tunnel.
Redis brokers the run ID and per-request telemetry between the load generator and the controller, and captures all run artifacts to disk.

\subsection{Backend Fleet and Fault Model}

Each backend derives its identity from its container hostname (reverse DNS, falling back to the hostname) and from that an instance number.
A mod-3 pattern over the instance number assigns profiles: pattern~0 is fast/reliable (8 ms base delay, 3 ms jitter, no errors), pattern~1 is fast/reliable (7 ms, 2 ms, no errors), and pattern~2, every third instance, is the designated faulty tier (9 ms, 4 ms, 5\% error rate); a small index-dependent delay term avoids exact duplicates as the fleet grows.
Behavior is deterministic per request: each request seeds a private RNG from a hash of the base seed, the instance, and the request ID, so latency and error draws are fully reproducible for a fixed seed.
Response delay is the base delay plus uniform jitter plus any spike, and errored requests return HTTP~500 with a small extra delay so failures do not dominate the latency statistics.
A no-error mode globally zeroes the error rate and is used only for the static baseline arm (the sole place the error-on versus error-off condition is varied).

\subsection{Telemetry and Control Surface}

Metrics come from HAProxy's Prometheus exporter (per-server status, current sessions, weight, and HTTP-response counters) and from each backend's own metrics endpoint, with Prometheus tracking the fleet by DNS service discovery.
Reads use the HAProxy Data Plane API (for backend and server discovery) and writes use the raw admin stats socket (for weight, drain, and enable commands).
Each control step issues Prometheus range queries over a 180 s lookback at 15 s resolution with a 2 m rate window for status, sessions, weight, and 5xx rate, assembling a unified snapshot of global aggregates (total sessions, total 5xx, servers up versus total) and a per-server list carrying current values plus recent series.
Each control step then proceeds observe--decide--act: the policy maps the snapshot to a plan of actions, the guardrail layer filters the plan, the surviving actions execute, and after a 0.5 s settle delay the controller re-observes and records the post-action state.
The guardrails enforce the per-step caps defined in the Methodology independent of what the policy proposed, and additionally clamp any executed weight to HAProxy's native $[0,256]$ range.
LLM policies additionally log the full per-step conversation and feed executed actions back into the next prompt.
The action vocabulary is the four function-calling tools (\texttt{set\_weight}, \texttt{drain}, \texttt{enable}, \texttt{no\_op}).

\subsection{LLM Serving}

The model is served through \texttt{llama.cpp}'s OpenAI-compatible chat-completions API, accessed via the standard OpenAI client, with full GPU offload, flash attention, the model's own chat template applied, a single inference thread, and a fixed decode seed (42); it runs either as a local subprocess or, for the full sweep, on a dedicated GPU host (a single NVIDIA RTX Pro 6000 with 96 GB of VRAM, rented via RunPod) reached over an SSH tunnel.
The registry spans the 15 models across the five families of Table~\ref{tab:models}, all served in their native (BF16/F16) precision.
Each control step uses a two-turn protocol; Appendix~\ref{app:control-loop-example} walks through a complete example step.
The first turn sends the system prompt (objective, observation format, balance algorithm, the server admin-state machine, and the action rules) plus a user message (recent action history and the formatted observation), with the four tools offered and tool choice left automatic; the second turn fires only if the model called tools, appending the assistant's tool-call message and per-call validation results (queued or error) with tool calls disabled, to elicit a final natural-language rationale.
Standard sampling parameters pass through the API directly, while the runtime's own knobs (top-$k$, min-$p$, repetition penalty, and the chat-template reasoning flags) pass through the client's extension field, following the per-family settings of Table~\ref{tab:sampling}.
Family-specific post-processing strips reasoning and tool-call markup, promoting reasoning into a unified field; full conversations, per-call validation errors, and per-turn token usage are logged.

\subsection{System Prompt}

The system prompt provides the LLM inside the controller with its objective, constraints, and interpretation rules for the observations.
It is instantiated at each control step with the active load-balancing algorithm (from the environment).
The complete prompt is:

\begin{lstlisting}[breaklines=true, basicstyle=\ttfamily\small]
You are the controller for an HAProxy load balancer in a closed loop. You are called once per interval, and each observation already reflects earlier actions.
Objective:
  Eliminate 5xx errors across the pool.
  Any server with 5xx_rate > 0 is a problem - shift traffic away from it.
HAProxy balance algorithm: {lb_algorithm}
Observation format:
  Global line:
    sessions    = total active connections across all servers
    5xx_per_sec = pool-wide 5xx errors per second
    servers_up  = healthy servers / total servers
  Per-server table columns:
    up       = 1 if server is healthy, 0 if down
    weight   = traffic share (1-100)
    sessions = active connections on this server
    load%    = this server's share of total sessions
    5xx_rate = HTTP 5xx errors per second
  Time series: shown only for metrics that are changing.
    RISING or FALLING appended when the trend is significant.
Server admin state is inferred from the last action taken per server:
  no prior action / enable / set_weight -> ACTIVE : valid actions = set_weight, drain
  drain -> DRAIN  : valid actions = enable
Action rule:
  If a server has 5xx_rate > 0 and it persists: call set_weight to halve its current weight.
  Do not drop weight below 1 - use drain instead to remove a server fully.
  When a server's 5xx_rate returns to 0 and stays at 0 for several intervals: gradually restore weight.
\end{lstlisting}

\subsection{Load Generation}

Load is generated by Grafana k6 (v1.4.2; see Section~\ref{sec:rw-k6} for its experimental pedigree) issuing \texttt{GET /work} through HAProxy, under either a steady profile (constant virtual users for a fixed duration, used throughout) or a ramping arrival-rate profile.
A per-virtual-user seeded PRNG drives a uniform inter-request sleep in $[5,50]$ ms, so identical seeds reproduce the arrival pattern.
A run ID is minted at setup and shared via Redis, so the controller picks up the same run without explicit handoff.
Per-request telemetry, timestamp, virtual user, iteration, status, duration, success flag, and the DNS/connect/TLS/time-to-first-byte/download timing breakdown are pushed to Redis.
The run summary aggregates k6's built-in metrics (request rate, request duration, including p99, and failure rate) for export to the run's output directory.

\section{Results}

All numbers in this section are computed from the 240-run corpus.
Fault-localization, stability, and guardrail statistics are computed over the 224 LLM runs, all of which execute with the error profile active (the static arms never propose a plan).
The 5xx-reduction figures compare each LLM run to the static, error-on run at the same scale and algorithm.

\subsection{Capability Threshold}

\begin{figure}[t]
  \centering
  \includegraphics[width=\linewidth]{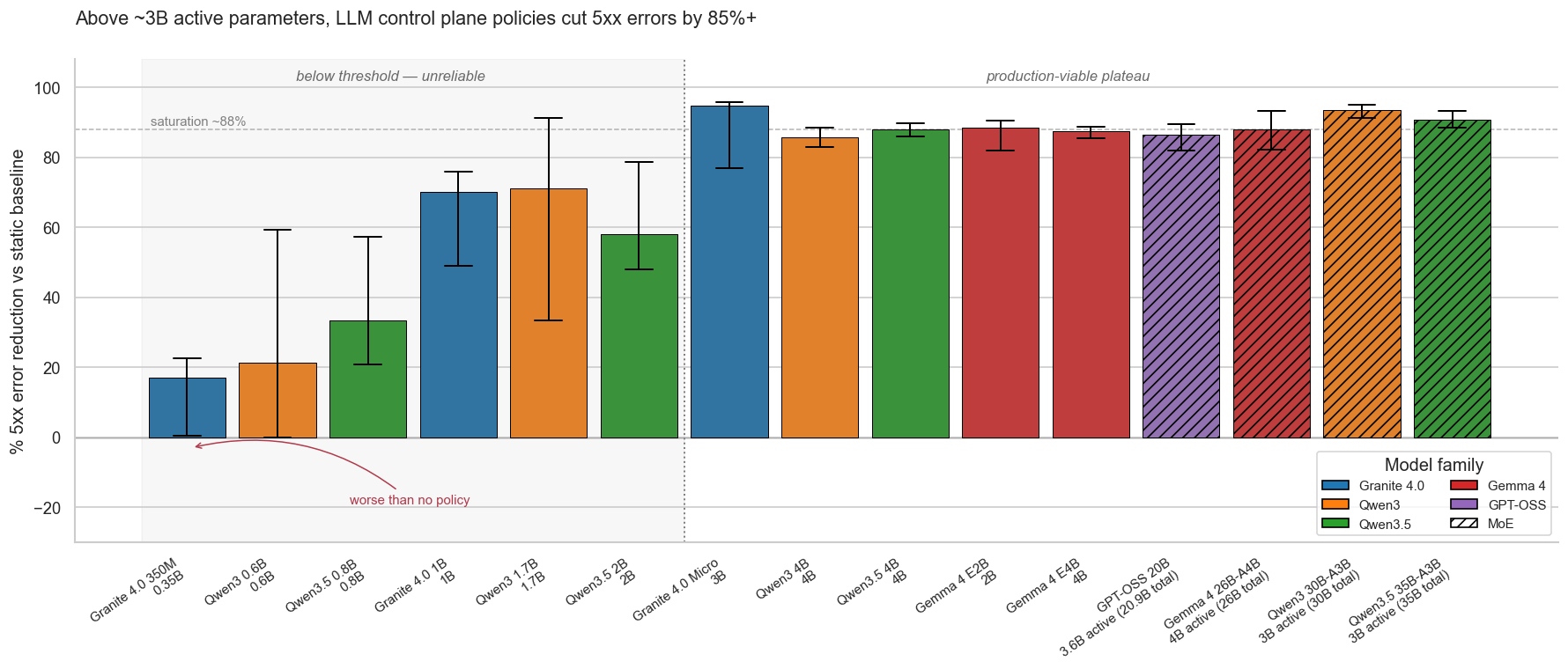}
  \caption{Median client-side 5xx reduction versus the matched static baseline, one bar per model, ordered by total then active parameters.
  Bars are colored by family and hatched for MoE checkpoints; black caps are 95\% bootstrap CIs across the scale $\times$ LB $\times$ thinking conditions.
  Below the capability threshold (shaded), models are unreliable and can do worse than no policy at all; above it, every architecture (dense, efficient-sparse, and MoE alike) saturates on an ${\sim}88\%$ plateau.}
  \label{fig:hero}
\end{figure}

The central result is a \emph{capability threshold} near 3 B active parameters, as shown in Figure~\ref{fig:hero} and observed across all metrics we measured.
Below the threshold, behavior is erratic: the smallest model (Granite~4.0 350M) never drains a server in any of its eight runs, and several small dense checkpoints either barely act or over-react, with the worst single configuration (Qwen3.5~2B with thinking, \texttt{leastconn}, $N{=}3$) regressing the client 5xx rate by $-667\%$ relative to doing nothing, i.e., sending nearly eight times as many failures to clients.
Above the threshold, the picture collapses onto a single plateau: Qwen3~4B, Qwen3.5~4B, all three Gemma~4 variants, GPT-OSS~20B, and both MoE Qwen checkpoints all land within a few points of an $\sim\!88\%$ median 5xx reduction, with dense, efficient-sparse, and MoE architectures statistically indistinguishable on the plateau.
The same boundary governs fault localization (Table~\ref{tab:loc-model}): micro-F1 jumps from $\le 0.81$ below the line to $\ge 0.87$ above it.
The boundary is approximate rather than a sharp parameter law, and two checkpoints straddle it: Gemma~4 E2B reaches the plateau with only 2 B active (5 B total) parameters, while the dense 3 B Granite~4.0 Micro does not.
We nevertheless use ``3 B active'' throughout as the closest single-number summary of where the regime changes.

\subsection{Static Baseline Characterization}

\begin{figure}[t]
  \centering
  \includegraphics[width=\linewidth]{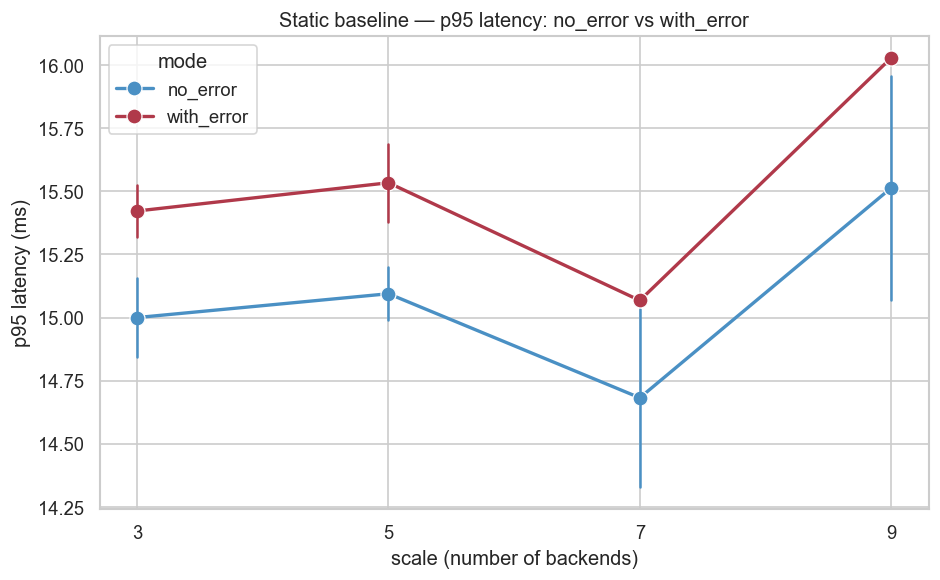}
  \caption{Static (uncontrolled) baseline, \texttt{no\_error} versus \texttt{with\_error} mode across scales, averaged over both LB algorithms.
  The error-on arm sits at the structural no-intervention floor; the no-error arm is the upper-bound reference an effective controller should approach.}
  \label{fig:static}
\end{figure}

The static baselines set the bar an LLM must clear.
Under no intervention, the aggregate failure rate is pinned by construction to the no-intervention floor of $1.4$--$2.0\%$ (measured $0.0141$--$0.0198$ across the four scales), and crucially \texttt{leastconn} does \emph{not} suppress the structural signal on its own: with connections closed per request, HAProxy's passive load-awareness routes the same proportional share to the lossy backends, so both schedulers sit at essentially the same floor.
Figure~\ref{fig:static} confirms this: the error-on static arm carries the full floor under both algorithms.
At the same time, the no-error reference (every backend at 0\% error) marks the latency-and-availability target a perfect controller should approach.

\subsection{Effectiveness Across Model Size and Family}

Pooled over the 224 LLM runs, fault localization is strong on average: micro precision/recall/F1 reach $0.913 / 0.792 / 0.848$, and exact set-match (predicted drained set $=$ faulty set) holds on $66.5\%$ of runs.
The pooled average, however, conceals a bimodal distribution across the threshold.
Table~\ref{tab:loc-model} breaks localization down by model in ascending size.
Everything at or below Granite~4.0 Micro is unreliable (F1 $\le 0.81$, often through one of two failure modes; see below), while every model from Qwen3~4B upward is saturated (F1 $\ge 0.87$).
Only Qwen3.5~4B and Gemma~4 E4B are perfect ($P{=}R{=}\text{exact}{=}1.0$); Qwen3~4B, Gemma~4 E2B, and the Gemma~4 26B-A4B MoE are near-perfect (F1 $0.984$, exact $0.938$).
Effectiveness medians tell the same story by family (Table~\ref{tab:eff-family}): every family that fields a supra-threshold checkpoint reaches $85$--$91\%$ median reduction, whereas Granite (whose largest member is the 3 B Micro) tops out at $70\%$.
The sub-threshold Qwen3 nothink configurations drag that family's nothink median down to $68\%$.

\begin{table}[t]
\centering
\caption{Fault localization by model, in ascending total-parameter order (224 LLM runs).
Micro scores pool TP/FP/FN; exact-match is the fraction of runs whose ever-drained set equals the faulty set.
The rule above Qwen3~4B marks the capability threshold.
Granite~4.0 350M never drains a server, so its precision is undefined.}
\label{tab:loc-model}
\small
\begin{tabular}{lrrrrr}
\hline
\textbf{Model} & \textbf{$n$} & \textbf{$P_\mu$} & \textbf{$R_\mu$} & \textbf{F1$_\mu$} & \textbf{Exact} \\
\hline
Granite 4.0 350M  &  8 & ---   & 0.000 & 0.000 & 0.000 \\
Qwen3 0.6B        & 16 & 0.917 & 0.344 & 0.500 & 0.188 \\
Qwen3.5 0.8B      & 16 & 0.690 & 0.625 & 0.656 & 0.250 \\
Granite 4.0 1B    &  8 & 0.692 & 0.562 & 0.621 & 0.000 \\
Qwen3 1.7B        & 16 & 1.000 & 0.500 & 0.667 & 0.500 \\
Qwen3.5 2B        & 16 & 0.714 & 0.938 & 0.811 & 0.375 \\
Granite 4.0 Micro &  8 & 0.667 & 1.000 & 0.800 & 0.500 \\
\hline
Qwen3 4B          & 16 & 1.000 & 0.969 & 0.984 & 0.938 \\
Qwen3.5 4B        & 16 & 1.000 & 1.000 & 1.000 & 1.000 \\
Gemma 4 E2B       & 16 & 1.000 & 0.969 & 0.984 & 0.938 \\
Gemma 4 E4B       & 16 & 1.000 & 1.000 & 1.000 & 1.000 \\
GPT-OSS 20B       & 24 & 1.000 & 0.854 & 0.921 & 0.875 \\
Gemma 4 26B-A4B   & 16 & 1.000 & 0.969 & 0.984 & 0.938 \\
Qwen3 30B-A3B     & 16 & 1.000 & 0.781 & 0.877 & 0.688 \\
Qwen3.5 35B-A3B   & 16 & 1.000 & 0.938 & 0.968 & 0.938 \\
\hline
\end{tabular}
\end{table}

\begin{table}[t]
\centering
\caption{Median client-side 5xx reduction (\%) versus the matched static baseline, by family and reasoning mode.
``---'' marks a combination not run (GPT-OSS has no non-reasoning mode; Granite has no reasoning support).
Cells dominated by sub-threshold members (Granite, Qwen3 nothink) sit well below the plateau.}
\label{tab:eff-family}
\small
\begin{tabular}{lrr}
\hline
\textbf{Family} & \textbf{nothink} & \textbf{think} \\
\hline
Gemma 4   & 90.9 & 86.0 \\
GPT-OSS   & ---  & 86.5 \\
Granite   & 70.0 & ---  \\
Qwen3     & 68.0 & 88.8 \\
Qwen3.5   & 85.1 & 85.4 \\
\hline
\end{tabular}
\end{table}

\subsubsection{Behavioral Regimes Below the Threshold}

Below the threshold, failure splits into two distinct behavioral regimes.
The \emph{conservative} (non-drainer) models drain rarely but correctly: Qwen3~0.6B acts with precision $0.92$ but recall only $0.34$, Qwen3~1.7B with precision $1.0$ but recall $0.50$, and Granite~4.0 350M never drains at all.
The \emph{trigger-happy} (over-drainer) models catch the faults but spray collateral onto healthy slots: Qwen3.5~2B (precision $0.71$, recall $0.94$) and Granite~4.0 Micro (precision $0.67$, recall $1.0$) are the only two models where recall exceeds precision.
This asymmetry is the corpus-wide dominant failure mode: pooled over all runs there are $355$ true positives against $93$ false negatives but only $34$ false positives, so faulty slots left in rotation outnumber healthy slots wrongly drained by roughly $3{\times}$.
The controller under-reacts far more often than it over-reacts.

A second, orthogonal failure is \emph{instability}, and it bites even the competent models.
Treating the final drained set as the controller's settled decision, end-state correctness is only $0.304$ across all runs, against an ever-drained exact-match of $0.665$: $83$ of the $149$ runs that at some point localized the fault correctly ($55.7\%$) had reverted that decision by the final step.
The churn lands squarely on the faulty slot: the macro faulty-slot flip rate is $4.12$ versus a healthy-slot flip rate of $0.066$, a $62{\times}$ gap.
$80.4\%$ of runs exhibit at least one oscillating server despite the system prompt's explicit hysteresis guidance, which, as worded, does not prevent flapping.

\subsubsection{Guardrail Firing Patterns}

The guardrail layer is not cosmetic: it is what keeps the unstable mid-tier and the reasoning-on configurations from issuing disruptive plans.
Because the processed logs preserve the model's \emph{raw} proposed plan alongside the executed one, we can measure exactly which proposals were illegal (Table~\ref{tab:guardrail-scale}, \S\ref{sec:rq2-guardrail}).
The rails fire on $1.63\%$ of all decision steps overall, but firing is highly concentrated: Qwen3.5~0.8B ($5.2\%$ of steps), the Qwen3~30B-A3B MoE ($4.0\%$), and Qwen3.5~2B ($3.3\%$) trip them most, while the perfect large models (Qwen3.5~35B-A3B, Gemma~4 E4B) and the inert Granite~4.0 350M never trip a rail at all.
The two dominant violations are weight clamps (a single $\pm50$-bounded step proposing a larger swing; $120$ steps, concentrated in the small Qwen3.5 models that lurch weights) and the drain cap (plans proposing two or three simultaneous drains; $62$ steps, concentrated in the Qwen3 thinking configurations).
Without the one-drain-per-cycle cap, those plans would quarantine multiple backends at once, precisely the catastrophic over-reaction that produced the worst regressions.

\subsection{Client Latency Across Control Arms}

\begin{figure}[t]
  \centering
  \includegraphics[width=\linewidth]{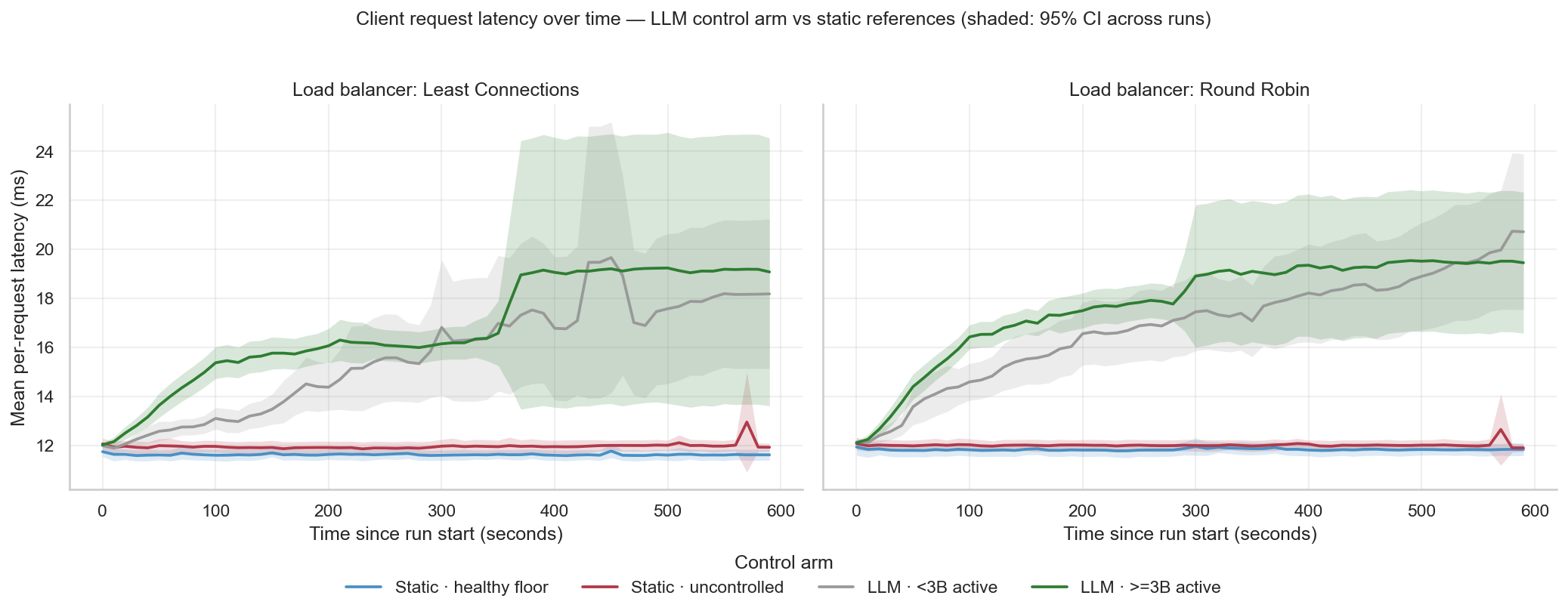}
  \caption{Client request latency over time, LLM control arms versus static references (shaded: 95\% CI on the mean), split by load-balancing algorithm and averaged over scales.
  Both static arms hold a flat ${\sim}12$ ms mean, while both LLM arms drift upward over the run as faulty backends are drained and traffic concentrates onto the surviving servers; the supra-threshold ($\ge$3 B) arm rises fastest.}
  \label{fig:latency}
\end{figure}

The same controller that suppresses 5xx errors also reshapes the client latency distribution (Figure~\ref{fig:latency}, Table~\ref{tab:latency}).
For the latency and token analyses, the LLM runs are split into two arms at exactly 3 B active parameters; note that this parameter cut places Gemma~4 E2B (2 B active) in the sub-threshold arm and Granite~4.0 Micro (3 B) in the supra-threshold arm, the two borderline checkpoints discussed above.
Both static arms hold a flat mean for the entire run: $11.77$ ms for the healthy floor and $12.00$ ms for the uncontrolled error-on baseline, with tight tails ($p95$ of $15.1$ and $15.6$ ms, $p99$ of $16.6$ and $17.7$ ms).
Both LLM arms instead drift upward over the $600$ s run: the supra-threshold ($\ge$3 B active) arm rises fastest, climbing from $\sim$12 ms at the start to $\sim$17 ms under \texttt{leastconn} and $\sim$18.6 ms under \texttt{roundrobin} by run end, with the sub-threshold ($<$3 B) arm tracking a few milliseconds below for most of the run.
Pooled across scales and schedulers, the supra-threshold arm's mean is $16.505$ ms (95\% CI $[16.503, 16.508]$, over $128$ runs and $103$ M requests) against $15.148$ ms for the sub-threshold arm.
The shift is concentrated in the tail: median latency is essentially unchanged across all four arms ($p50$ of $11.5$--$12.1$ ms), while the supra-threshold arm's $p95$ and $p99$ ($40.4$ and $49.0$ ms) are roughly $2.6{\times}$ and $2.8{\times}$ the uncontrolled baseline.
Relative to the references, the supra-threshold controller adds $+4.74$ ms mean, $+25.3$ ms $p95$, and $+32.4$ ms $p99$ versus the healthy floor, and $+4.50$, $+24.8$, and $+31.3$ ms versus the uncontrolled baseline.

The tail penalty grows with fleet size (Table~\ref{tab:latency-scale}).
The supra-threshold arm's $p95$ rises from $26$--$28$ ms at $N{=}3$ to $45$--$47$ ms at $N{=}9$, overtaking the sub-threshold arm as the fleet grows: at $N{=}3$ the supra-threshold tail is the lower of the two ($26.1$/$27.6$ ms for leastconn/roundrobin versus $33.0$/$37.6$ ms), but by $N{=}9$ it is the higher ($45.3$/$47.2$ ms versus $30.9$/$45.0$ ms).
Both static arms remain flat near $15$--$16$ ms $p95$ across all scales.

\begin{figure}[t]
  \centering
  \includegraphics[width=\linewidth]{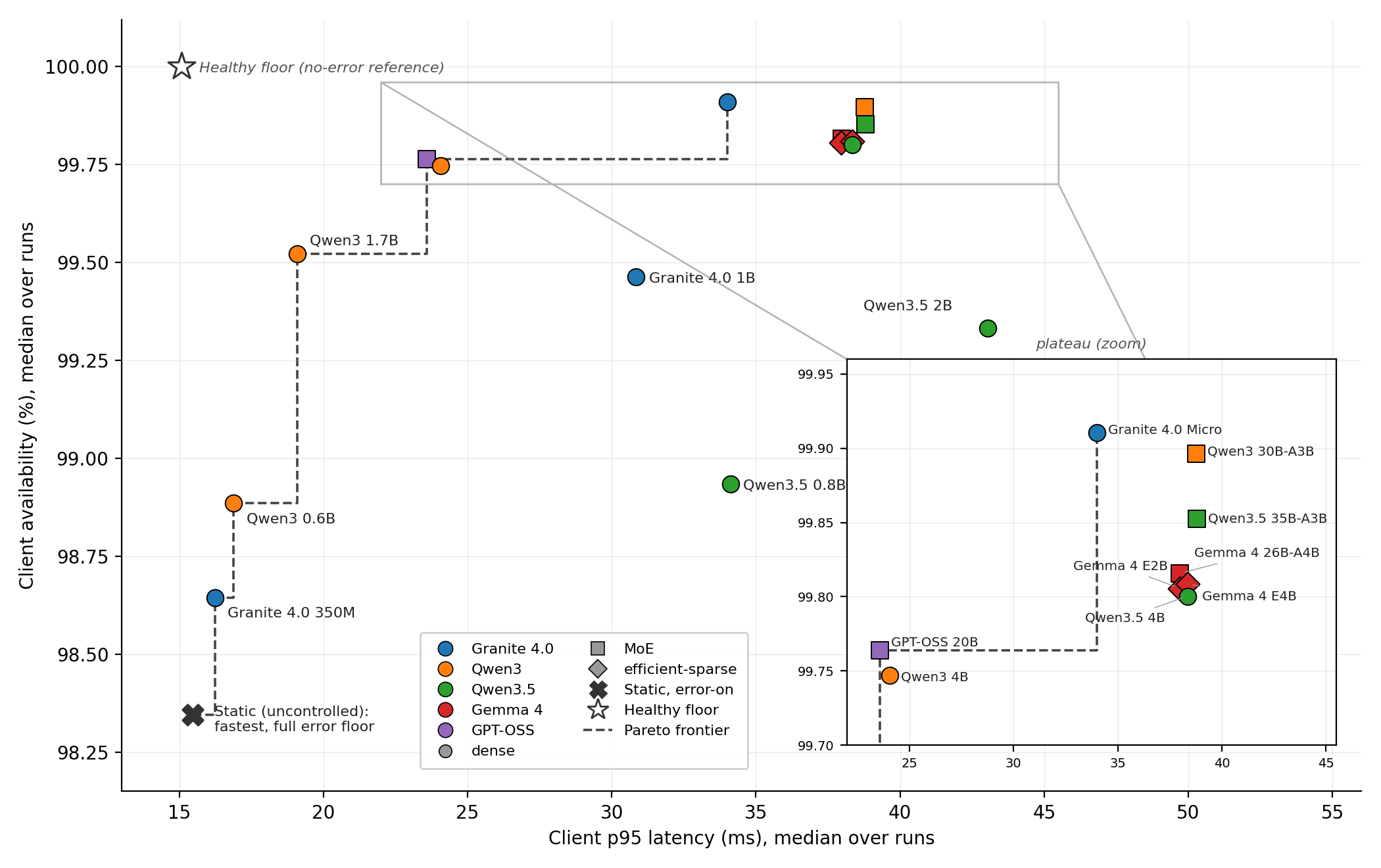}
  \caption{Availability--latency Pareto view.
  One marker per model: median client availability ($100\times(1-\text{5xx rate})$) versus median client $p95$ latency over that model's error-on runs (scale $\times$ LB $\times$ reasoning mode).
  Color encodes family as in Figure~\ref{fig:hero}; marker shape encodes architecture; every point is labeled directly.
  The error-on static baseline ($\times$) competes as a do-nothing policy and anchors the fast-but-unavailable end of the dashed Pareto frontier; the no-error static arm ($\star$) is the unreachable healthy-floor reference.
  The inset zooms the plateau cluster.
  Control buys ${\sim}1.4$--$1.6$ points of availability for a $1.5$--$2.5\times$ higher $p95$ tail, with GPT-OSS~20B at the knee of the frontier.}
  \label{fig:pareto}
\end{figure}

Figure~\ref{fig:pareto} recasts the two client-side outcomes as an availability--latency Pareto view: one point per model (median client availability, $100\times(1-\text{5xx rate})$, against median client $p95$ latency, pooled over that model's error-on runs), with the error-on static baseline competing as a do-nothing policy and the no-error arm as the unreachable healthy-floor reference.
The frontier makes the price of control explicit: relative to doing nothing, the supra-threshold models buy back roughly $1.4$--$1.6$ percentage points of client availability at a $1.5$--$2.5\times$ higher $p95$ tail.
GPT-OSS~20B sits at the knee of the frontier, delivering plateau-level availability at ${\sim}24$ ms $p95$, while most plateau models cluster near $38$ ms.
Granite~4.0 Micro's frontier position is instructive rather than contradictory: its characteristic over-draining (localization precision $0.667$, exact-match $0.5$) maximizes the availability axis precisely because it quarantines aggressively, so the axis rewards it even though its decision quality is borderline; availability alone therefore under-reports the difference between it and the cleanly-localizing plateau models.

\begin{table}[t]
\centering
\caption{Client request latency by control arm, pooled across scales and LB algorithms.
Means carry a 95\% CI; percentiles are over all requests in the arm.
The two LLM arms group the error-on runs by a strict 3 B active-parameter cut; the two static arms are the healthy-floor (no-error) and uncontrolled (error-on) references.}
\label{tab:latency}
\small
\begin{tabular}{lrrrrrr}
\hline
\textbf{Arm} & \textbf{Runs} & \textbf{Requests} & \textbf{Mean (ms)} & \textbf{$p50$} & \textbf{$p95$} & \textbf{$p99$} \\
\hline
Static $\cdot$ healthy floor & 8   & 7{,}223{,}335   & 11.767 & 11.5 & 15.1 & 16.6 \\
Static $\cdot$ uncontrolled  & 8   & 7{,}033{,}886   & 12.001 & 11.9 & 15.6 & 17.7 \\
LLM $\cdot$ $<$3B active      & 96  & 79{,}432{,}077  & 15.148 & 12.1 & 38.3 & 47.9 \\
LLM $\cdot$ $\ge$3B active    & 128 & 103{,}238{,}823 & 16.505 & 12.1 & 40.4 & 49.0 \\
\hline
\end{tabular}
\end{table}

\begin{table}[t]
\centering
\caption{Client latency $p95$ (ms) by fleet size $N$, LB algorithm, and arm (error-on arms).
The supra-threshold arm's tail overtakes the sub-threshold arm as $N$ grows, while the uncontrolled static reference stays flat.}
\label{tab:latency-scale}
\small
\begin{tabular}{rlrrr}
\hline
\textbf{$N$} & \textbf{LB algo} & \textbf{LLM $<$3B} & \textbf{LLM $\ge$3B} & \textbf{Static uncontrolled} \\
\hline
3 & leastconn  & 33.0 & 26.1 & 15.3 \\
  & roundrobin & 37.6 & 27.6 & 15.5 \\
5 & leastconn  & 38.7 & 39.0 & 15.4 \\
  & roundrobin & 39.2 & 40.6 & 15.6 \\
7 & leastconn  & 31.2 & 35.7 & 15.1 \\
  & roundrobin & 38.2 & 36.7 & 15.1 \\
9 & leastconn  & 30.9 & 45.3 & 16.0 \\
  & roundrobin & 45.0 & 47.2 & 16.0 \\
\hline
\end{tabular}
\end{table}

\subsection{Scaling Behavior}

This subsection addresses RQ2: how the controller's detection quality, reaction time, effectiveness, and guardrail pressure change as the backend fleet grows.

\subsubsection{Fault Localization as the Fleet Grows}

As the fleet grows, the controller must detect more faulty slots in a larger universe, and recall pays for it (Table~\ref{tab:loc-scale}).
Recall trends down from $0.893$ at $N{=}3$ to $0.732$ at $N{=}9$ (non-monotone: it recovers at $N{=}7$ before falling at $N{=}9$).
Precision moves in the opposite direction, from $0.877$ to $0.961$: at larger scale, the controller misses more of the now-more-numerous faults but stays disciplined about not touching healthy servers.
Exact-match is highest at the easiest single-fault scale ($0.786$ at $N{=}3$) and lowest where all three faults must be caught together ($0.589$ at $N{=}9$).

\begin{table}[t]
\centering
\caption{Fault localization and guardrail firing by fleet size $N$ (56 LLM runs each).
Recall falls, and precision rises with scale; exact set-match is hardest, where all faults must be caught.
The final column is the share of decision steps on which any guardrail fired.}
\label{tab:loc-scale}
\small
\begin{tabular}{rrrrrr}
\hline
\textbf{$N$} & \textbf{$P_\mu$} & \textbf{$R_\mu$} & \textbf{F1$_\mu$} & \textbf{Exact} & \textbf{Guard.\ fire} \\
\hline
3 & 0.877 & 0.893 & 0.885 & 0.786 & 0.56\% \\
5 & 0.873 & 0.795 & 0.832 & 0.589 & 1.79\% \\
7 & 0.912 & 0.830 & 0.869 & 0.696 & 1.82\% \\
9 & 0.961 & 0.732 & 0.831 & 0.589 & 2.38\% \\
\hline
\end{tabular}
\end{table}

\subsubsection{Time-to-Action and Time-to-Resolution}

Reaction time stretches with the fleet (Table~\ref{tab:resolution}).
Median time-to-action (the first drain of \emph{any} faulty server) stays in the $60$--$120$ s band across scales (6--12 decision intervals).
Time-to-resolution (every faulty server drained at least once) climbs steeply instead, because there are more servers to cover: from $59$--$79$ s at $N{=}3$ (one fault, so action and resolution coincide) to $240$--$289$ s at $N{=}9$ (three faults).
Across the corpus, $195$ of the $224$ LLM runs ($87\%$) eventually drain at least one faulty server and $168$ ($75\%$) drain all of them, so full coverage is the common case but not universal, and it arrives later as the fleet grows.

\begin{table}[t]
\centering
\caption{Median time-to-action and time-to-resolution (seconds from run start) by fleet size and LB algorithm.
Time-to-action is the first drain of any faulty server; time-to-resolution is when all faulty servers have been drained at least once.
At $N{=}3$ the single fault makes the two coincide.}
\label{tab:resolution}
\small
\begin{tabular}{llrr}
\hline
\textbf{$N$} & \textbf{LB algo} & \textbf{Time-to-action (s)} & \textbf{Time-to-resolution (s)} \\
\hline
3 & leastconn  &  79.0 &  79.0 \\
  & roundrobin &  59.0 &  59.0 \\
5 & leastconn  &  84.0 & 152.0 \\
  & roundrobin & 107.0 & 178.0 \\
7 & leastconn  &  63.5 & 171.0 \\
  & roundrobin &  64.0 & 170.0 \\
9 & leastconn  & 108.0 & 288.5 \\
  & roundrobin & 116.5 & 240.0 \\
\hline
\end{tabular}
\end{table}

\subsubsection{Error Reduction at Each Fleet Scale}

Effectiveness erodes gracefully with scale rather than collapsing (Table~\ref{tab:eff-scale}).
Median 5xx reduction is highest at $N{=}3$ ($89.6\%$/$93.2\%$ for leastconn/roundrobin) and falls to $77.4\%$/$76.9\%$ at $N{=}9$, still a more-than-three-quarters reduction at the largest fleet.
The two schedulers track each other closely at every scale.
The erosion is consistent with the recall story above: at $N{=}9$ the controller is slower to cover all three faults (Table~\ref{tab:resolution}) and more likely to leave one in rotation.
Hence, a residual slice of the floor survives to the client.

\begin{table}[t]
\centering
\caption{Median client-side 5xx reduction (\%) versus the matched static baseline, by fleet size and LB algorithm.
Effectiveness degrades gracefully with scale and is near-identical across schedulers.}
\label{tab:eff-scale}
\small
\begin{tabular}{rrr}
\hline
\textbf{$N$} & \textbf{leastconn} & \textbf{roundrobin} \\
\hline
3 & 89.6 & 93.2 \\
5 & 85.2 & 84.1 \\
7 & 86.4 & 89.9 \\
9 & 77.4 & 76.9 \\
\hline
\end{tabular}
\end{table}

\subsubsection{Guardrail Firing Rate by Fleet Size}
\label{sec:rq2-guardrail}

\begin{figure}[t]
  \centering
  \includegraphics[width=\linewidth]{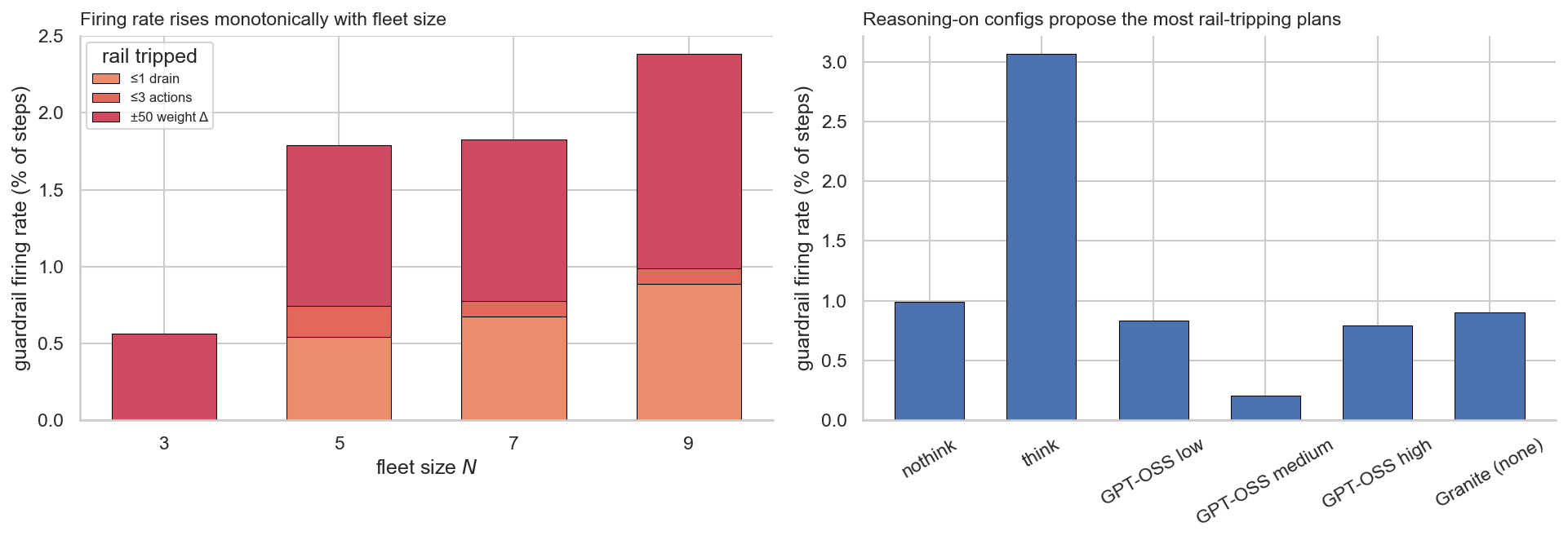}
  \caption{Guardrail firing.
  \emph{Left:} firing rate by fleet size, stacked by rail; the rate climbs monotonically with $N$, and the drain cap (dormant at $N{=}3$) grows as models try to quarantine several backends at once.
  \emph{Right:} firing rate by reasoning configuration; thinking-on and the small Qwen3.5 dense models propose the most rail-tripping plans, while the perfect large models and the inert Granite~350M never trip a rail.}
  \label{fig:guardrail}
\end{figure}

Guardrail firing rises monotonically with the fleet, from $0.56\%$ of steps at $N{=}3$ to $2.38\%$ at $N{=}9$ (Table~\ref{tab:guardrail-scale}, Figure~\ref{fig:guardrail}).
The composition shifts with scale as well: at $N{=}3$ the only firings are weight clamps, because with a single fault there is never a reason to propose two drains, whereas from $N{=}5$ upward the drain cap begins to fire ($16$, $20$, and $26$ steps at $N{=}5,7,9$) as models attempt to quarantine several faulty backends in a single cycle.
In other words, the larger the fleet, the more often the model's instinct is to act more aggressively than the safety rails permit.

\begin{table}[t]
\centering
\caption{Guardrail firing by fleet size $N$, decomposed by rail.
A step ``fires'' if the raw model plan proposes more than one drain (drain cap), more than three actions (action cap), or a single weight change exceeding $\pm50$ (weight clamp).
No proposal ever requested an out-of-range weight.
The drain cap is dormant at $N{=}3$ (a single fault) and grows with the fleet.}
\label{tab:guardrail-scale}
\small
\begin{tabular}{rrrrrr}
\hline
\textbf{$N$} & \textbf{Steps} & \textbf{Drain cap} & \textbf{Action cap} & \textbf{Weight clamp} & \textbf{Any (\%)} \\
\hline
3 & 3029 &  0 & 0 & 17 & 0.56 \\
5 & 2964 & 16 & 6 & 31 & 1.79 \\
7 & 2960 & 20 & 3 & 31 & 1.82 \\
9 & 2936 & 26 & 3 & 41 & 2.38 \\
\hline
\end{tabular}
\end{table}

\subsection{Reasoning Effort: Quality and Cost}

This subsection addresses RQ3, separating what reasoning does for decision quality from what it costs the control loop, at fixed model weights.

\subsubsection{Decision Quality With and Without Reasoning}

Within a fixed checkpoint, turning reasoning on does not move effectiveness: median 5xx reduction is $86.6\%$ for nothink configurations versus $86.3\%$ for think configurations (Table~\ref{tab:budget}), and the same near-parity holds in the family medians of Table~\ref{tab:eff-family}.
For GPT-OSS, the three reasoning-effort levels actually invert the intuition: low and medium effort reach $88.5\%$ and $88.1\%$, but \emph{high} effort drops to $74.0\%$.
Reasoning does buy a modest improvement in the \emph{settled} state: thinking-on runs end correctly $38.4\%$ of the time versus $22.3\%$ for thinking-off.
However, it buys that improvement by flapping \emph{more} along the way (mean $8.64$ versus $6.07$ flips per run).
On the primary client-facing metric, reasoning is close to free of effect.

\subsubsection{Inference Cost and Control-Loop Cadence}

The cost side of that ledger is steep (Table~\ref{tab:budget}, Figure~\ref{fig:token}).
Enabling thinking multiplies median completion tokens per step from $157$ to $1{,}546$ ($\sim$10$\times$) and per-run completion tokens from $18.5$ k to $76.6$ k; GPT-OSS at high effort emits $3{,}428$ completion tokens \emph{per step} and $111$ k per run.
Because the controller runs on a 10 s cycle, this cost converts directly into missed deadlines: nothink, low/medium GPT-OSS, and Granite all hold a $10$ s median cadence and complete the full $60$ decision steps, but thinking-on slips to a $12$ s median cadence with $72.7\%$ of runs falling behind real time and only $\sim$50 of 60 steps completed, and GPT-OSS-high slips to an $18$ s cadence with \emph{every} run falling behind and a median of only $27.5$ of $60$ steps completed.

\begin{table}[t]
\centering
\caption{Inference budget by reasoning configuration (224 LLM runs).
\emph{Comp/step} is median completion tokens per step; \emph{Comp/run} is mean total completion tokens.
\emph{Cadence} is the median inter-step wall-clock; values above 10 s mean the control loop is slipping behind real time (a derived proxy: the inter-step delta covers the whole observe$\to$infer$\to$execute cycle, not isolated Inference).
\emph{Steps} is the median number of decision steps completed of a nominal 60.
\emph{Over budget} is the share of runs whose median cadence exceeds 10 s.
\emph{Reduct.}\ is the median \% 5xx reduction.
\emph{Cost-eff}\ is the cost-effectiveness ratio: failed requests avoided (absolute 5xx reduction versus the matched static baseline) per $1{,}000$ completion tokens spent over the run.}
\label{tab:budget}
\small
\resizebox{\linewidth}{!}{%
\begin{tabular}{lrrrrrr}
\hline
\textbf{Metric} & \textbf{nothink} & \textbf{think} & \textbf{GPT-OSS} & \textbf{GPT-OSS} & \textbf{GPT-OSS} & \textbf{Granite} \\
 & & & \textbf{low} & \textbf{med} & \textbf{high} & \textbf{(none)} \\
\hline
Comp/step & 157 & 1546 & 166 & 689 & 3428 & 87 \\
Comp/run & 18480 & 76618 & 10183 & 45206 & 110954 & 5187 \\
Cadence (s) & 10.0 & 12.0 & 10.0 & 10.0 & 18.0 & 10.0 \\
Steps & 60.0 & 50.0 & 60.0 & 60.0 & 27.5 & 60.0 \\
Over budget & 0\% & 73\% & 0\% & 0\% & 100\% & 0\% \\
Reduct. & 86.6 & 86.3 & 88.5 & 88.1 & 74.0 & 70.0 \\
Cost-eff & 762 & 131 & 1214 & 280 & 93 & 1138 \\
\hline
\end{tabular}%
}
\end{table}

\subsubsection{Cost-Effectiveness}

\begin{figure}[t]
  \centering
  \includegraphics[width=\linewidth]{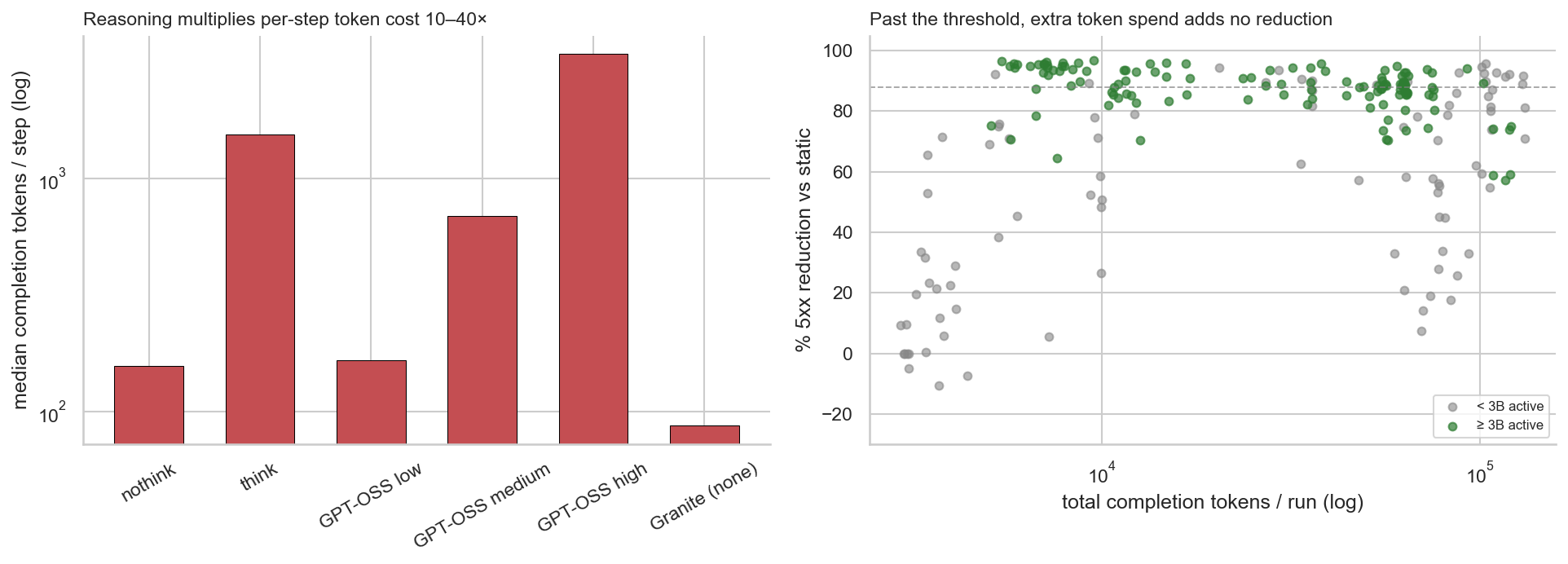}
  \caption{Inference cost.
  \emph{Left:} median completion tokens per step by reasoning configuration (log scale); thinking mode multiplies per-step cost ${\sim}10\times$, and GPT-OSS high effort ${\sim}21\times$ over low.
  \emph{Right:} cost-effectiveness, per-run completion tokens (log) versus \% 5xx reduction, colored by whether the model clears the 3 B active-parameter threshold.
  Above the threshold (green), the reduction plateau is flat over two orders of magnitude in token spend; below it (grey), additional spend does not lift the model out of the unreliable regime.}
  \label{fig:token}
\end{figure}

Combining the two halves, the cost-effectiveness frontier is flat above the capability threshold (Figure~\ref{fig:token}, right).
Once a model clears $\sim$3 B active parameters, every additional order of magnitude of completion-token spend buys essentially no extra 5xx reduction: the supra-threshold runs cluster in the $85$--$97\%$ band whether they spent $5$ k or $110$ k completion tokens.
Expressed as a return-on-spend ratio (failed requests avoided per $1{,}000$ completion tokens, Table~\ref{tab:budget}), the non-reasoning configurations dominate: GPT-OSS at low effort returns $1{,}214$ and the nothink configurations $762$, against $131$ for thinking-on, $280$ for GPT-OSS medium, and $93$ for GPT-OSS high, a $6$--$13{\times}$ gap at near-equal 5xx reduction.
Granite posts a comparably high ratio ($1{,}138$) on the smallest token budget of any family ($5.2$ k completion tokens per run), despite its lower $70\%$ absolute reduction.

\section{Discussion}

We now interpret the results against the three research questions in turn, then bound the claims with the threats to their validity.

\subsection{Conditions for Effective LLM Control}

Addressing RQ1, an LLM control plane reduces client-perceived 5xx errors below the static baseline once the controller clears a capability threshold of roughly 3 B active parameters; below it, the same pattern is unreliable and can underperform doing nothing.
Against this floor, a competent LLM controller removes the overwhelming majority of client-visible failures: the best single configurations reach $+96$--$97\%$ reduction (Qwen3~4B, \texttt{roundrobin}, $N{=}3$: $+96.9\%$, 95\% Newcombe CI $[+94.6, +99.2]$), and the within-run Wilson intervals on each run's 5xx rate are a fraction of a percentage point wide, so the spread \emph{across} conditions in Figure~\ref{fig:hero} is a real design effect rather than sampling noise.

The sub-threshold failures split along two axes.
The first is a conservative-versus-trigger-happy divide in how readily a model drains: the conservative models drain rarely but correctly.
In contrast, the trigger-happy ones catch the faults but spray collateral onto healthy slots.
The second, orthogonal failure is instability that bites even competent models, with the majority of the runs that at some point localized the fault correctly reverting that decision by the final step.
The instability is the controller second-guessing a correct drain (consistent with the high precision above), not randomly thrashing healthy capacity, which is why the system prompt's hysteresis guidance, as worded, does not prevent the flapping.
This mirrors IntentContinuum's use of an LLM as a real-time decision-maker \cite{11169635}, but at the load-balancer layer rather than the orchestration layer; the present results show the same LLM-as-controller pattern works one layer down, provided its action space is bounded.

The guardrails therefore function as a deterministic safety floor that makes the sub-perfect models deployable rather than merely a tidiness measure: without the one-drain-per-cycle cap, the rail-tripping plans would quarantine multiple backends at once, precisely the catastrophic over-reaction that produced the worst regressions.

However, this gain in effectiveness is not free, and the latency results qualify the RQ1 answer.
Because the controller concentrates traffic onto the healthy majority by draining the lossy backends, it removes capacity from the pool.
The surviving servers absorb the same offered load at higher utilization, so the supra-threshold arm pays roughly $+4.7$ ms in mean latency and a $\sim$2.6--2.8$\times$ inflation of the $p95$/$p99$ tail relative to either static reference (Table~\ref{tab:latency}), even as it eliminates the overwhelming majority of 5xx errors.
The median request is barely affected; the cost lands in the tail and grows with the drained fraction, which is why the $p95$ climbs toward $45$ ms at $N{=}9$, where a full third of the fleet is quarantined.
Whether this is a good trade is an SLO question, and the Pareto view of Figure~\ref{fig:pareto} states its terms: the controller is strictly preferable when availability dominates the objective, but a latency-sensitive SLO would favor softer de-weighting (partial weight reduction rather than full drains) to keep capacity online.
The practical condition for effective control is thus twofold: a supra-threshold model, and a guardrail layer that makes its occasional over-reactions safe.

\subsection{Scalability of the Closed Loop}

Addressing RQ2, effectiveness degrades gracefully rather than catastrophically as the fleet grows, from $89$--$93\%$ at $N{=}3$ to roughly $77\%$ at $N{=}9$, still more than a three-quarters reduction at the largest fleet.
At larger scale, the controller misses more of the now-more-numerous faults but stays disciplined about not touching healthy servers, so the degradation is a \emph{recall} degradation: the system grows conservative, not reckless, as it scales.
The close tracking of the two schedulers at every scale confirms that the controller, not the underlying HAProxy algorithm, drives the reduction.
The composition of guardrail firings also shifts with scale, the drain cap activating only once multiple faults coexist, which is a quantitative argument that the guardrails become more load-bearing precisely as the problem gets harder.

Because the bottleneck at scale is recall (faults left in rotation rather than healthy servers wrongly drained), the multi-agent IBN frameworks \cite{jiang2026agenticaiempoweredintentbased,11131150} suggest a natural remedy: decomposing the fleet across specialist controllers could raise recall by giving each agent a smaller universe to cover, at the cost of the inter-agent credit-assignment problem those works highlight \cite{10.5555/3709347.3743784}.
For deployments larger than the nine-backend fleet studied here, improving recall (whether by agent decomposition or by a longer observation window) is the priority, not tighter precision.
Scale also sharpens the capacity-versus-availability trade: draining a larger subset of faulty servers removes more from the pool.
Hence, the tail-latency cost of control rises with $N$ (Table~\ref{tab:latency-scale}), the supra-threshold arm's $p95$ overtaking the sub-threshold arm by $N{=}9$ precisely because it drains more aggressively.

\subsection{Reasoning as a Control-Loop Resource}

Addressing RQ3, reasoning effort does not improve the primary client-facing metric: median 5xx reduction is $86.6\%$ without reasoning versus $86.3\%$ with it, a difference within noise.
The reasoning trace appears to help the model eventually re-converge after extra intermediate oscillation rather than to make crisper first decisions.
The dropped steps are the mechanism behind the regression in GPT-OSS-high's effectiveness.
A controller that completes only $27.5$ of $60$ intervals observes the fleet less than half as often.
Hence, it detects faults later and re-evaluates its quarantine decisions less frequently, exactly the conditions under which a residual error slice survives to the client.
The reasoning effort that should make each decision better instead degrades the closed loop by starving it of decisions: the $74.0\%$ median reduction at high effort sits \emph{below} the $88.5\%$/$88.1\%$ of the cheaper low/medium settings on the same model and the same hardware.

This is the clearest practical finding for deployment: under a fixed control cadence, reasoning budget must be capped so that the decision lands inside the window.
The efficient operating point is therefore a supra-threshold model run in its \emph{cheapest} mode: GPT-OSS at low effort ($10$ k completion tokens/run, $88.5\%$ reduction) or a nothink 4 B dense model, both of which match or beat their own reasoning-on variants at a fraction of the token and latency cost.
The return-on-spend ratio makes this concrete: the cheapest non-reasoning configurations avoid $6$--$13{\times}$ more failed requests per token than their reasoning-on counterparts (e.g.\ GPT-OSS low at $1{,}214$ versus high at $93$), so for a fixed inference budget the dominant lever is turning reasoning \emph{off}, not scaling the model up.
However, the ratio must be read with caution: Granite's high score ($1{,}138$) reflects its tiny token budget rather than strong control, since its absolute reduction is capped at $70\%$.
The metric rewards frugality and should be paired with the absolute reduction, not substituted for it.
Spending more inference time only helps below the threshold, where no amount of test-time compute can rescue a sub-3 B model, and even there, the gains are erratic rather than reliable.
This contrasts with the reasoning models that produce near-optimal economic dispatch schedules in prior work \cite{11259643}: those are open-loop, single-shot problems where additional test-time compute is unambiguously beneficial, whereas a fixed control cadence imposes a hard wall on inference time, turning extra reasoning into missed decisions.

\subsection{Threats to Validity}

Three standard categories of threat bound the claims above.

\subsubsection{Construct Validity}

Our primary effectiveness metric is the client-side 5xx reduction, which we treat as a proxy for SLO satisfaction; it captures availability but not the full latency or fairness dimensions a production SLO might encode.
Fault localization depends on a post hoc identification of the faulty set (the slots with non-zero cumulative 5xx counts); the labeling separated cleanly in every run, but it presumes a fault that manifests in the counters at all.
Finally, the controller observes only server-side metrics while effectiveness is judged client-side, leaving an observability gap between what the model sees and what the user experiences.

\subsubsection{Internal Validity}

All runs share a single seed family (seed 42 throughout), so the reported draws are reproducible but do not characterize run-to-run variance under reseeding.
The system prompt was fixed and not ablated, so we cannot separate the model's contribution from that of the prompt's wording, including its hysteresis guidance.
The controller poll interval was held fixed at 10 s and not varied, so the interaction between cadence and effectiveness, central to the RQ3 findings, is observed at a single operating point.

\subsubsection{External Validity}

The evaluation uses a single fault class, one load balancer (HAProxy), one backend stack (Flask), and one telemetry source (Prometheus), so generalization to other fault patterns and stacks is untested.
The fleet never exceeds nine backends, leaving behavior at production scale open.
Inference is self-hosted on a single GPU (an NVIDIA RTX Pro 6000, 96 GB) with no per-token pricing and no production network latency; the token counts are hardware-agnostic, but the cadence and steps-completed figures reflect this GPU's throughput and would shift on faster or slower serving hardware.
The model set is closed and the serving stack pinned to \texttt{llama.cpp}, so results may shift with other inference runtimes or newer checkpoints.

\section{Conclusion}

Static load balancers cannot mitigate a backend that is degraded rather than down, leaving the resulting 5xx burden on client-perceived availability.
We framed this as an intent-assurance problem and replaced HAProxy's static routing policy with a closed-loop LLM controller that observes per-backend server-side metrics every 10 seconds and adjusts server weights and drain state through guardrailed tool calls.
Across 240 runs covering 15 open-weight models (with reasoning toggled or graded where each model supports it), two scheduling algorithms, and four fleet scales against a deterministic persistent-fault benchmark, we measured when this control plane beats the static baseline, how it scales, and what reasoning effort buys.

The results turn on a capability threshold near 3 B active parameters.
With respect to RQ1, models above it remove the overwhelming majority of client-visible 5xx errors (an $\sim$88\% median reduction, peaking at $+96$--$97\%$).
In contrast, the models below it are unreliable and can do worse than no policy at all.
This availability gain carries a latency cost: draining the lossy backends concentrates load onto fewer servers, raising mean latency by ${\sim}4.7$  ms and inflating the $p95$/$p99$ tail by ${\sim}2.6$--$2.8\times$ over the static references.
The controller is therefore a clear win only when availability dominates the objective.
With respect to RQ2, effectiveness degrades gracefully as the fleet grows, from $89$--$93\%$ at $N{=}3$ to roughly $77\%$ at $N{=}9$.
The degradation is a recall problem: the controller grows conservative rather than reckless.
With respect to RQ3, reasoning effort does not improve the primary client-facing metric and, when it overruns the control interval, actively degrades it by starving the loop of decisions; on a per-token basis, the cheapest non-reasoning configurations avoid $6$--$13\times$ more failed requests than their reasoning-on counterparts.

The practical recommendation follows directly: where availability is the dominant objective, run a supra-threshold model in its cheapest, non-reasoning mode, and treat the action-plan guardrails not as a tidiness measure but as a deterministic safety floor that makes otherwise-unstable models deployable.
Where tail latency is also constrained, the same controller should prefer softer de-weighting over full drains, trading a little availability headroom to keep capacity online.

Several directions remain open.
The benchmark exercises a single persistent fault class, so extending it to transient, gradual-ramp, and cascading fault patterns would test the controller under conditions it was deliberately spared here.
Broader infrastructure coverage (other load balancers, backend stacks, and telemetry sources, and fleets well beyond nine backends) would establish how far the threshold and the recall-limited scaling behavior generalize.
Policy learning and distillation could compress a supra-threshold controller into a cheaper model, and cost-aware control could make the reasoning budget itself a decision variable tied to the loop cadence.
Finally, situating the load-balancer controller within the full IBN lifecycle, from specification and translation through to this assurance loop, would close the gap between operator intent and runtime control that motivated the work.

\clearpage

\appendix

\section{Model Weight Provenance}

\label{app:checkpoints}

\begin{table}[h!]
\centering
\caption{Weight provenance for full reproducibility.
Each model is the GGUF artifact at the listed HuggingFace repository and commit.
Commit is the 7-character HuggingFace revision; ``Pulled'' is the download date.}
\label{tab:checkpoints}
\small
\begin{tabular}{llll}
\hline
\textbf{HuggingFace repo} & \textbf{Quant} & \textbf{Commit} & \textbf{Pulled} \\
\hline
unsloth/Qwen3-0.6B-GGUF         & BF16 & 50968a4 & 2025-06-23 \\
unsloth/Qwen3-1.7B-GGUF         & BF16 & d7f544e & 2025-06-08 \\
unsloth/Qwen3-4B-GGUF           & BF16 & 22c9fc8 & 2025-06-08 \\
unsloth/Qwen3-30B-A3B-GGUF      & BF16 & d5b1d57 & 2025-06-08 \\
\hline
unsloth/Qwen3.5-0.8B-GGUF       & BF16 & 7fc504f & 2026-03-02 \\
unsloth/Qwen3.5-2B-GGUF         & BF16 & e16cafb & 2026-03-02 \\
unsloth/Qwen3.5-4B-GGUF         & BF16 & b5044a2 & 2026-03-02 \\
unsloth/Qwen3.5-35B-A3B-GGUF    & BF16 & 4b0737a & 2026-03-03 \\
\hline
unsloth/gemma-4-E2B-it-GGUF     & BF16 & fd2ea40 & 2026-04-11 \\
unsloth/gemma-4-E4B-it-GGUF     & BF16 & 8543410 & 2026-04-11 \\
unsloth/gemma-4-26B-A4B-it-GGUF & BF16 & b19ae87 & 2026-05-04 \\
\hline
ibm-granite/granite-4.0-350m-GGUF  & BF16 & 2954f7a & 2025-10-28 \\
ibm-granite/granite-4.0-1b-GGUF    & BF16 & 9309b02 & 2025-12-18 \\
ibm-granite/granite-4.0-micro-GGUF & F16  & 397e2dc & 2025-10-07 \\
\hline
unsloth/gpt-oss-20b-GGUF        & F16  & d449b42 & 2025-12-19 \\
\hline
\end{tabular}
\end{table}

\section{Example Control Loop Interaction}

\label{app:control-loop-example}

This appendix shows a complete example of one control step from the closed-loop system, illustrating the two-turn protocol, observation format, tool calls, and model reasoning.

\subsection{Input: Recent Action History}

\begin{lstlisting}[breaklines=true, basicstyle=\ttfamily\small]
Recent actions (oldest -> newest):
  180s ago be_api/api-3 set_weight w=50
  120s ago be_api/api-1 drain
\end{lstlisting}

\subsection{Input: Observation}

\begin{lstlisting}[breaklines=true, basicstyle=\ttfamily\small]
Global sessions=287 5xx_per_sec=2.34 servers_up=3/4

Backend be_api
server up weight sessions load% 5xx_rate
api-1 0 0 0 0% 0.0
api-2 1 100 94 33% 0.8
api-3 1 50 95 33% 1.2 trend=-0.3/60s FALLING
api-4 1 100 98 34% 0.3

Time series (oldest to newest, 15s intervals):
api-3 5xx_rate=[1.5,1.2,1.2]
api-4 5xx_rate=[0.1,0.2,0.3] trend=+0.2/30s RISING
\end{lstlisting}

\subsection{Turn 1: Tool Calls}

The model receives the system prompt and user message (above), and is offered four tools.
It makes the following calls:

\begin{lstlisting}[breaklines=true, basicstyle=\ttfamily\small]
Tool call 1: set_weight
  backend: "be_api"
  server: "api-3"
  weight: 25
  reason: "5xx_rate=1.2/s persisting; halving weight from 50"

Tool call 2: set_weight
  backend: "be_api"
  server: "api-4"
  weight: 75
  reason: "5xx_rate rising to 0.3/s; reducing weight to prevent escalation"
\end{lstlisting}

\subsection{Turn 2: Tool Validation and Reasoning}

The server validates both tool calls and returns status ``queued'' for each.
The model then provides a natural-language rationale:

\begin{lstlisting}[breaklines=true, basicstyle=\ttfamily\small]
The pool is under load with 287 active sessions. Two servers are showing
elevated 5xx rates: api-3 at 1.2/s (falling but still significant) and
api-4 at 0.3/s (rising trend). api-1 remains drained from the prior
action. By halving api-3's weight and reducing api-4, we shift traffic
toward the healthy api-2 (0.8/s) and reduce exposure to error-prone
servers. The falling trend on api-3 suggests recovery; api-4 needs
early intervention given the rising slope.
\end{lstlisting}

\subsection{Logged Conversation}

The complete interaction (system message, user message, tool calls, tool results, and reasoning) is stored as a JSON entry in the conversation log, along with per-turn token usage and action validation results.
This enables post-hoc analysis of decision quality, reasoning patterns, and inference cost across the sweep.

\bibliographystyle{elsarticle-num}

\bibliography{references}

@INPROCEEDINGS{11259643,
  author={Mohammadi, Sina and Hassan, Ali and Haghighi, Rouzbeh and Bui, Van-Hai and Su, Wencong},
  booktitle={2025 IEEE Energy Conversion Conference Congress and Exposition (ECCE)}, 
  title={Large Language Models for Solving Economic Dispatch Problem}, 
  year={2025},
  volume={},
  number={},
  pages={1-5},
  doi={10.1109/ECCE58356.2025.11259643}}

@INPROCEEDINGS{10760725,
  author={Bokkena, Bhargava},
  booktitle={2024 2nd International Conference on Self Sustainable Artificial Intelligence Systems (ICSSAS)}, 
  title={Optimizing Cloud Infrastructure Management Using Large Language Models: A DevOps Perspective}, 
  year={2024},
  volume={},
  number={},
  pages={1401-1406},
  doi={10.1109/ICSSAS64001.2024.10760725}}

@INPROCEEDINGS{11308679,
  author={Tan, Yanyan and Zhang, Wei and Tan, Lizhuang and Zhang, Peiying and Zhang, Wei and Sun, Jiande},
  booktitle={2025 International Conference on New Trends in Computational Intelligence (NTCI)}, 
  title={Collaborative Operator Selection for MultiObjective Optimization via Large Language Models and Deep Reinforcement Learning}, 
  year={2025},
  volume={},
  number={},
  pages={1-5},
  doi={10.1109/NTCI67886.2025.11308679}}

@INPROCEEDINGS{11101733,
  author={Abuaziz, Ahmed and Celiktas, Baris},
  booktitle={2025 9th International Symposium on Innovative Approaches in Smart Technologies (ISAS)}, 
  title={A Context-Aware, AI-Driven Load Balancing Framework for Incident Escalation in SOCs}, 
  year={2025},
  volume={},
  number={},
  pages={1-10},
  doi={10.1109/ISAS66241.2025.11101733}}

@ARTICLE{11293797,
  author={Hossain, Md. Kamrul and Aljoby, Walid},
  journal={IEEE Open Journal of the Communications Society}, 
  title={NetIntent: Leveraging Large Language Models for End-to-End Intent-Based SDN Automation}, 
  year={2025},
  volume={6},
  number={},
  pages={10512-10541},
  doi={10.1109/OJCOMS.2025.3642642}}

@INPROCEEDINGS{11311632,
  author={Islam, Koushikur and Da Cunha Rodrigues, Guilherme and Javadi, Bahman and Calheiros, Rodrigo N.},
  booktitle={2025 IEEE International Conference on Smart Internet of Things (SmartIoT)}, 
  title={MicroIntent: Intent-Based Placement Strategy for Microservice Application in the Compute Continuum Using LLMs}, 
  year={2025},
  volume={},
  number={},
  pages={124-131},
  doi={10.1109/SmartIoT66867.2025.00026}}

@INPROCEEDINGS{10539172,
  author={Manias, Dimitrios Michael and Chouman, Ali and Shami, Abdallah},
  booktitle={2024 20th International Conference on the Design of Reliable Communication Networks (DRCN)}, 
  title={Towards Intent-Based Network Management: Large Language Models for Intent Extraction in 5G Core Networks}, 
  year={2024},
  volume={},
  number={},
  pages={1-6},
  doi={10.1109/DRCN60692.2024.10539172}}

@inproceedings{10.5555/3709347.3743784,
author = {Nagpal, Kartik and Dong, Dayi and Mehr, Negar},
title = {Leveraging Large Language Models for Effective and Explainable Multi-Agent Credit Assignment},
year = {2025},
isbn = {9798400714269},
publisher = {International Foundation for Autonomous Agents and Multiagent Systems},
address = {Richland, SC},
booktitle = {Proceedings of the 24th International Conference on Autonomous Agents and Multiagent Systems},
pages = {1501–1510},
numpages = {10},
location = {Detroit, MI, USA},
series = {AAMAS '25}
}

@INPROCEEDINGS{11169635,
  author={Akbari, Negin and Grundy, John and Cheema, Aamir and Toosi, Adel N.},
  booktitle={2025 IEEE International Conference on Web Services (ICWS)}, 
  title={IntentContinuum: Using LLMs to Support Intent-Based Computing Across the Compute Continuum}, 
  year={2025},
  volume={},
  number={},
  pages={573-583},
  doi={10.1109/ICWS67624.2025.00079}}

@misc{jiang2026agenticaiempoweredintentbased,
      title={Agentic AI Empowered Intent-Based Networking for 6G}, 
      author={Genze Jiang and Kezhi Wang and Xiaomin Chen and Yizhou Huang},
      year={2026},
      eprint={2601.06640},
      archivePrefix={arXiv},
      primaryClass={cs.AI},
      url={https://arxiv.org/abs/2601.06640}, 
}

@inproceedings{10.1145/3651890.3672268,
author = {Wu, Duo and Wang, Xianda and Qiao, Yaqi and Wang, Zhi and Jiang, Junchen and Cui, Shuguang and Wang, Fangxin},
title = {NetLLM: Adapting Large Language Models for Networking},
year = {2024},
isbn = {9798400706141},
publisher = {Association for Computing Machinery},
address = {New York, NY, USA},
url = {https://doi.org/10.1145/3651890.3672268},
doi = {10.1145/3651890.3672268},
booktitle = {Proceedings of the ACM SIGCOMM 2024 Conference},
pages = {661–678},
numpages = {18},
location = {Sydney, NSW, Australia},
series = {ACM SIGCOMM '24}
}

@ARTICLE{10418089,
  author={Dzeparoska, Kristina and Tizghadam, Ali and Leon-Garcia, Alberto},
  journal={IEEE Communications Magazine}, 
  title={Emergence: An Intent Fulfillment System}, 
  year={2024},
  volume={62},
  number={6},
  pages={36-41},
  doi={10.1109/MCOM.001.2300270}}

@ARTICLE{11131150,
  author={Brodimas, Dimitrios and Birbas, Alexios and Kapolos, Dimitrios and Denazis, Spyros},
  journal={IEEE Open Journal of the Communications Society}, 
  title={Intent-Based Infrastructure and Service Orchestration Using Agentic-AI}, 
  year={2025},
  volume={6},
  number={},
  pages={7150-7168},
  doi={10.1109/OJCOMS.2025.3600706}}

@INPROCEEDINGS{11186656,
  author={Tageldien, Marwa and Selim, Bassant and Sboui, Lokman},
  booktitle={2025 International Telecommunications Conference (ITC-Egypt)}, 
  title={Large Language Models in Intent-Based Networking: a Comprehensive Survey Across the Intent Lifecycle}, 
  year={2025},
  volume={},
  number={},
  pages={810-817},
  doi={10.1109/ITC-Egypt66095.2025.11186656}}

@INPROCEEDINGS{11125105,
  author={Gharbaoui, M. and Castoldi, P.},
  booktitle={2025 25th Anniversary International Conference on Transparent Optical Networks (ICTON)}, 
  title={Artificial Intelligence in Intent-Based Networking: Enhancing Automation, Scalability, and Reliability}, 
  year={2025},
  volume={},
  number={},
  pages={1-4},
  doi={10.1109/ICTON67126.2025.11125105}}

@INPROCEEDINGS{11073734,
  author={Dzeparoska, Kristina and Leon-Garcia, Alberto},
  booktitle={NOMS 2025-2025 IEEE Network Operations and Management Symposium}, 
  title={Intent-Based Management for Network Automation}, 
  year={2025},
  volume={},
  number={},
  pages={1-7},
  doi={10.1109/NOMS57970.2025.11073734}}

@ARTICLE{11003942,
  author={Angi, Antonino and Sacco, Alessio and Marchetto, Guido},
  journal={IEEE Transactions on Network and Service Management}, 
  title={LLNet: An Intent-Driven Approach to Instructing Softwarized Network Devices Using a Small Language Model}, 
  year={2025},
  volume={22},
  number={4},
  pages={3403-3418},
  doi={10.1109/TNSM.2025.3570017}}

@INPROCEEDINGS{11073607,
  author={Nam, Sukhyun and Van Tu, Nguyen and Hong, James Won-Ki},
  booktitle={NOMS 2025-2025 IEEE Network Operations and Management Symposium}, 
  title={LLM-Based AI Agent for VNF Deployment in OpenStack Environment}, 
  year={2025},
  volume={},
  number={},
  pages={1-7},
  doi={10.1109/NOMS57970.2025.11073607}}

@ARTICLE{10742582,
  author={Njah, Yosra and Leivadeas, Aris and Falkner, Matthias},
  journal={IEEE Communications Magazine}, 
  title={An AI-Driven Intent-Based Network Architecture}, 
  year={2025},
  volume={63},
  number={4},
  pages={146-153},
  doi={10.1109/MCOM.001.2400143}}

@ARTICLE{11297203,
  author={Zhu, Longlong and Yu, Jiashuo and Chen, Xiang and Jiang, Zhifan and Liu, Xuan and Zhang, Jianshan and Yang, Xu and Zhang, Ruichen and Niyato, Dusit and Yi, Xun and Khalil, Ibrahim and Zhang, Dong and Wu, Chunming},
  journal={IEEE Communications Magazine}, 
  title={Leveraging Large Language Models for Multi-Objective and Adaptive SFC Deployment: Techniques, Case Study, and Promising Directions}, 
  year={2026},
  volume={64},
  number={1},
  pages={32-39},
  doi={10.1109/MCOM.001.2500254}}

@article{https://doi.org/10.1002/spe.70000,
author = {Aqasizade, Hossein and Ataie, Ehsan and Bastam, Mostafa},
title = {Kubernetes in Action: Exploring the Performance of Kubernetes Distributions in the Cloud},
journal = {Software: Practice and Experience},
volume = {55},
number = {10},
pages = {1711-1725},
doi = {https://doi.org/10.1002/spe.70000},
url = {https://onlinelibrary.wiley.com/doi/abs/10.1002/spe.70000},
eprint = {https://onlinelibrary.wiley.com/doi/pdf/10.1002/spe.70000},
year = {2025}
}

@Article{app152412991,
AUTHOR = {Giurgică, Teodor-Mihail and Sârbu, Annamaria and Klauer, Bernd and Găină, Liviu},
TITLE = {Field-Deployable Kubernetes Cluster for Enhanced Computing Capabilities in Remote Environments},
JOURNAL = {Applied Sciences},
VOLUME = {15},
YEAR = {2025},
NUMBER = {24},
ARTICLE-NUMBER = {12991},
URL = {https://www.mdpi.com/2076-3417/15/24/12991},
ISSN = {2076-3417},
DOI = {10.3390/app152412991}
}

@InProceedings{10.1007/978-3-031-99854-6_11,
author="Gu, Jianfeng
and Wang, Puxuan
and David N{\'u}{\~{n}}ez Araya, Isaac
and Huang, Kai
and Gerndt, Michael",
editor="Nagel, Wolfgang E.
and Goehringer, Diana
and Diniz, Pedro C.",
title="HAS-GPU: Efficient Hybrid Auto-scaling with Fine-Grained GPU Allocation for SLO-Aware Serverless Inferences",
booktitle="Euro-Par 2025: Parallel Processing",
year="2026",
publisher="Springer Nature Switzerland",
address="Cham",
pages="159--174",
isbn="978-3-031-99854-6"
}

@article{hebbar2025sla,
  author    = {Hebbar, Kishore Subramanya},
  title     = {Priority-Aware Reactive {APIs}: Leveraging {Spring WebFlux} for
               {SLA}-Tiered Traffic in Financial Services},
  journal   = {European Journal of Electrical Engineering and Computer Science},
  year      = {2025},
  volume    = {9},
  number    = {5},
  pages     = {31--40},
  doi       = {10.24018/ejece.2025.9.5.743}
}

@InProceedings{10.1007/978-3-032-10344-4_3,
author="Pandey, Vaibhav",
editor="Barolli, Leonard
and Ishida, Tomoyuki
and Dantas, Mario",
title="Reinforcement Learning-Based Autoscaling for Cost and Performance Optimization in Kubernetes Clusters",
booktitle="Advances on P2P, Parallel, Grid, Cloud and Internet Computing",
year="2026",
publisher="Springer Nature Switzerland",
address="Cham",
pages="25--37",
isbn="978-3-032-10344-4"
}

\end{document}